\documentclass[pra,twocolumn,showpacs,floatfix]{revtex4}
\usepackage{graphicx}
\usepackage[usenames]{color} 
\newcommand{\beq}{\begin{equation}}
\newcommand{\eeq}{\end{equation}} 
\newcommand{\beqa}{\begin{eqnarray}}
\newcommand{\eeqa}{\end{eqnarray}} 
\newcommand{\ba}{\begin{array}}
\newcommand{\ea}{\end{array}}

\begin{document}

\title{ Symmetry breaking in a localized interacting binary
BEC in a bi-chromatic 
optical lattice}

\author{Yongshan Cheng$^{1,2}$\footnote{yong\_shan@163.com} 
and
   S. K. Adhikari$^1$\footnote{adhikari@ift.unesp.br;
URL: www.ift.unesp.br/users/adhikari}}
\affiliation{$^1$Instituto de F\'{\i}sica Te\'orica, UNESP - Universidade 
Estadual Paulista, 
01.140-070 S\~ao Paulo, S\~ao Paulo, Brazil\\
$^2$Department of Physics, Hubei Normal University, Huangshi 435002, China
}

\begin{abstract} By direct numerical simulation of the time-dependent 
Gross-Pitaevskii equation using the split-step Fourier spectral method 
we study different aspects of the localization of a cigar-shaped {\it 
interacting} binary (two-component) Bose-Einstein condensate (BEC) in a 
one-dimensional bi-chromatic quasi-periodic optical-lattice potential, 
as used in a recent experiment on the localization of a BEC [Roati {\it 
et al.}, Nature {\bf 453}, 895 (2008)]. We consider two types of 
localized states: (i) when both localized components have a maximum of 
density at the origin $x=0$, and (ii) when the first component has a 
maximum of density and the second a minimum of density at $x=0$. In the 
non-interacting case the density profiles are symmetric around $x=0$. We 
numerically study the breakdown of this symmetry due to inter-species 
and intra-species interaction acting on the two components.  Where 
possible, we have compared the numerical results with a time-dependent 
variational analysis. We also demonstrate the stability of the localized 
symmetry-broken BEC states under small perturbation.

\end{abstract}


\maketitle

\section{Introduction}

After the prediction of the localization of the electronic wave function 
in a disordered potential by Anderson fifty years ago \cite{anderson}, 
there have been many studies on different aspects of localization of 
different types of waves, e.g., electron wave, sound wave, 
electromagnetic wave, etc \cite{light,micro,sound}, 
and quantum matter wave in the form of a Bose-Einstein condensate (BEC)
\cite{billy,roati,chabe,edwards}. In the case of a quantum matter wave a 
disordered speckle potential \cite{billy} and a quasi-periodic 
bi-chromatic optical lattice (OL) potential \cite{roati} have been 
employed for the localization  of a non-interacting cigar-shaped
BEC. The quasi-periodic bi-chromatic OL potential was formed by the 
superposition of two standing-wave polarized laser beams with incommensurate
wavelengths. Each standing-wave polarized 
laser beam leads to a periodic OL potential. The 
non-interacting BEC of $^{39}$K atoms was created \cite{roati} by tuning 
the inter-atomic scattering length to zero near a Feshbach resonance 
\cite{fesh}.

The localization of a cigar-shaped BEC in a one-dimensional (1D)
bi-chromatic OL potential
and related topics have been the subject matter of
several theoretical
\cite{das,dnlse,boers,random,optical,modugno,roux,roscilde,paul,adhikari}
and experimental 
\cite{billy,roati,chabe,edwards}
studies. {{(Localized BEC states in the form of gap soliton in
a mono-chromatic OL potential have also been observed \cite{gsexp} and studied
\cite{gs}. However, 
there are fundamental differences between the two types of localized states.
The Anderson localized states in bi-chromatic OL potential appear predominantly 
in the linear Schr\"odiger equation and are destroyed above a small critical 
repulsive nonlinearity, whereas the gap solitons appear only in the presence 
of a repulsive nonlinearity. The linear system does not support a localized 
gap soliton. The gap solitons appear in the band gap of the spectrum of the 
periodic OL potential. On the other hand, the potential supporting the Anderson states must be 
aperiodic in nature.)}}
After the pioneering experiments [5, 6] on the
localization of a 1D cigar-shaped non-interacting BEC, a natural extension
of this phenomenon would be to investigate localization
in a weakly-interacting binary (two-component) BEC mixture.

In this paper, with intensive numerical simulations of the 
Gross-Pitaevskii (GP) equation, we study different aspects of 
localization {of a cigar-shaped binary BEC} in a 1D bi-chromatic quasi-periodic 
OL potential $-$ similar to the one used in the experiment of Roati {\it 
et al.} \cite{roati}. Although, most of the studies of Anderson 
localization were confined to non-interacting systems, in the present 
study on the localization of a two-component BEC we shall mostly 
consider a weakly-interacting system under the action of both 
inter-species and intra-species interaction. If the potential $V(x)$ is 
symmetric around origin at $x=0$, e.g. $V(x)=V(-x)$, 
for the non-interacting system the 
density profiles are symmetric around $x=0$: $|u(x)|^2=|u(-x)|^2,$ where 
$u(x)$ is the wave function of the matter wave. But for the interacting 
system with inter- and intra-species interaction, this symmetry can be 
broken and different types of symmetry-broken states emerge and we 
shall be specially interested in the study of the formation of localized 
cigar-shaped 
BEC components with broken symmetry.

The two components of the BEC could be two different hyperfine states of 
the same atom \cite{binary1}
or two different atoms \cite{binary2}
and the two confining bi-chromatic 
OL potential for the two components could be the same or different
\cite{adhikari,modugno,roati}. 
The 
bi-chromatic OL potential could be taken as the linear combination of two 
sine terms, both with a minimum at the origin ($x=0$), when 
the confined BEC 
components will have a maximum at the origin. The bi-chromatic OL 
potential could also be taken as the linear combination of two cosine 
terms with a maximum at the origin, while the confined BEC components 
will have a minimum at the origin. We shall consider two distinct 
situations of localized binary BEC: (i) two components under the 
action of the identical sine bi-chromatic OL potential while both 
densities have maxima at $x=0$, and (ii) component 1 under the action of 
the sine bi-chromatic OL potential with a maximum of density at $x=0$ and 
component 2 under the action of the cosine bi-chromatic OL potential with 
a minimum of density at $x=0$. In case (i) with weak inter-species 
and intra-species repulsions, the localized 
BEC components are overlapping with a maximum at $x=0$. 
But under the 
action of a inter-species repulsion above a critical value, the two repealing 
BEC 
components separate and move to opposite sides of 
$x=0$, thus breaking the symmetry around $x=0$. 
It is this separation or splitting of the BEC components which we 
study. (Similar separation of gap solitons \cite{gs} in a binary BEC mixture confined
by a monochromatic lattice has been studied \cite{malomed,malomed2}.)
In case (ii) a different type of symmetry breaking emerges. In this 
case with weak intra-species and inter-species repulsion or 
attraction, the localized BEC components have densities symmetric around 
$x=0$.  But under the action of an inter-species attraction above a 
critical value, the two attracting BEC components 
try to stick together as a pair 
of bright solitons \cite{cpld}
and the minimum-energy stable configuration is the one where 
both components move predominantly to the same side of $x=0$, thus breaking 
the symmetry around $x=0$. We study this symmetry breaking in details.
In our study we use direct numerical simulation of the underlying 
GP equation using the split-step Fourier spectral method. 
We also use a time-dependent 
variational analysis for an analytical understanding of the 
numerical results in case (i) above when both components have a maximum at 
$x=0$ with densities having  quasi-Gaussian shapes.  In this case certain 
aspects of the splitting of the two components are also predicted 
correctly from the variational analysis.


There have already been a number of { theoretical}  and experimental  
studies on Anderson localization 
in different problems. The effect of a repulsive 
non-linearity on localization of light waves in photonic
crystals was studied experimentally \cite{lahini}. There have also been 
studies of a weak repulsion on localization 
\cite{dnlse,modugno,adhikari,interaction}.
 Damski {\it et al.} and Schulte {\it et al.}
 considered Anderson localization in disordered OL
potential \cite{optical} whereas 
Sanchez-Palencia {\it et
al.} and Cl\'ement {\it et
al.} considered Anderson localization in a random potential
\cite{random}.  There have been studies of Anderson
localization with other types of disorder \cite{other}.
Anderson localization in BEC under the action of a disordered potential
in two and three dimensions has also been investigated \cite{2D3D}.

In Sec. \ref{II} we present a brief account of (i) the two-component 1D GP 
equation,  (ii) the bi-chromatic OL potentials used in our study and 
(iii) a 
time-dependent variational analysis of the GP equation  under appropriate 
conditions. In Sec. \ref{III} we present our numerical studies on 
localization using the split-step Fourier spectral method. The density 
profile of the quasi-Gaussian localized states are in agreement with the 
variational results. The density of the localized states are symmetric 
around the origin at $x=0$ in the absence of inter-species interaction. 
We study the symmetry breaking under the action of an inter-species 
interaction. Certain aspects of symmetry breaking can be explained with 
the variational analysis.
We demonstrate  the dynamical stability of the 
two-component localized states.   
In Sec. \ref{IIII} we present a brief discussion and 
concluding remarks.

\section{Analytical consideration of localization}

\label{II}

We consider a two-component cigar-shaped BEC under tight transverse harmonic
confinement with the bi-chromatic OL potential acting along 
the axial $x$ direction. Then it is appropriate to consider a 1D reduction 
\cite{1d}
of the 3D GP equation by freezing the transverse dynamics to the
respective ground state and integrating over the transverse variables. 
Such a binary BEC is 
described by the following coupled 1D BEC equation \cite{1d,malomed}:
\begin{eqnarray}\label{eq1}
i\frac{\partial u_1}{\partial t}=- 
\frac{1}{2}\frac{\partial^2 u_1}{\partial x^2}
+g_1|u_1|^2+g_{12}|u_2|^2 u_1+V_1(x)u_1,\\
i\frac{\partial u_2}{\partial t}=- \frac{1}{2}\frac{\partial^2 u_2}
{\partial x^2}
+g_2|u_2|^2+g_{12}|u_1|^2 u_2+V_2(x)u_2,\label{eq2}
\end{eqnarray}
with normalization $\int_{-\infty}^{\infty}|u_j|^2 dx=1,$ of the 
localized wave function $u_j$ of the two components $j=1,2$. As the 
localization of the BEC in the bi-chromatic lattice is most prominent for 
the linear problem, and we shall be interested in small non-linearities 
$g_1, g_2$ and $g_{12}$. To make the parameters of the 
model tractable we take the number of atoms $N_j$ and the mass $m_j$ of 
the two components to be equal: $N_1=N_2=N$ and $m_1=m_2=m$. This 
simplification will have no effect on the general conclusion of this study.
The intra-species nonlinearities are given by  \cite{1d}
$g_j= 2a_jN/a_\perp , a_\perp \equiv \sqrt{\hbar/(m\omega)}, 
\omega \equiv \sqrt{\omega_y\omega_z}$ where 
 $\omega_y$ and $\omega_z$ are the transverse trap frequencies 
in the $y$ and $z$ 
directions, respectively, and $a_j$ is the intra-species atomic scattering 
length. The inter-species non-linearity is given by \cite{1d} 
 $g_{12}=2Na_{12}/a_\perp$, where $a_{12}$ is the 
inter-species  scattering length. 
In
Eqs. (\ref{eq1}) and (\ref{eq2}) $V_j(x)$ are the  
bi-chromatic OL traps of the two components 
$j=1,2$. 
Here we are using harmonic oscillator units: length is expressed in units of 
transverse oscillator length $a_\perp $ 
and time in units of angular frequency 
$(\omega)^{-1}$.

In the first part of our study we shall take 
the two bi-chromatic OL potentials to be identical $-$  $V_1(x)=V_2(x)=V(x)$ 
$-$ 
and  have 
the following sine  form as in the 
experiment \cite{roati}
\begin{eqnarray}\label{pot1}
V(x)=\sum_{l=1}^2 A_l \sin^2(k_lx),
\end{eqnarray}
with $A_l=2\pi^2 s_l/\lambda_l^2$, where $\lambda_l$'s 
are the wavelengths of the laser forming the OL potentials, $s_l$'s are their 
intensities, and  $k_l=2\pi/\lambda_l$ the corresponding wave number.
In the second part of our study we shall take $V_1(x)$ to be given by 
Eq. (\ref{pot1}) whereas in  $V_2(x)$  the sine term is replaced 
by a cosine term:
\begin{eqnarray}\label{pot2}
V_2(x)=\sum_{l=1}^2 A_l \cos^2(k_lx),
\end{eqnarray} 
Potentials (\ref{pot1}) and (\ref{pot2}) are quite similar.
However,  potential (\ref{pot2}) generates a different type of
localized states
compared to potential  (\ref{pot1}). Potential  (\ref{pot1}) has a local
minimum at the center $x=0$;  consequently,  stable stationary solutions
with this potential have a maximum at $x=0$. However, potential
(\ref{pot2}) has a local maximum at $x=0$ corresponding to a minimum of
the stationary solution at the center.

With  a single periodic potential of the form $\sin^2(2 \pi x/\lambda)$
or $\cos^2(2 \pi x/\lambda)$ 
with $s_2=0,$  
the linear Schr\"odinger equation permits only 
de-localized states in the form of Bloch waves. Localization is possible 
in the linear Schr\"odinger equation due to the ``disorder'' introduced 
through a second component in Eqs. (\ref{pot1}) or (\ref{pot2}). 
The primary lattice  is usually strong enough and is used as the main 
periodic potential fixing the Bloch band structure of the 
single-particle states without disorder. The secondary lattice is weaker 
($s_2<<s_1$) and introduces a ``deterministic" disorder. 
Following the experiment of Roati {\it et al.} \cite{roati}, the transverse 
harmonic-oscillator length is of the order of a few microns so that the 
wavelengths of the OL in dimensionless units become 
approximately $10^0   \sim  10^1$.

Usually, the stationary BEC localized states 
formed with bi-chromatic lattices occupy many sites of the quasi-periodic 
OL potential and their shape is modulated by the short-wavelength 
potential \cite{bishop}. 
For
certain values of the parameters, potential (\ref{pot1}) leads to
bound states confined practically to the central site of the
quasi-periodic OL potential. When this happens, a variational
approximation
with Gaussian ansatz leads to a reasonable prediction for the bound
state.
To apply the variational approach with potential (\ref{pot1}) effective 
on both components, we adopt the Gaussian function below 
as the variational ansatz \cite{PG}
\begin{eqnarray}\label{va}
u_j(x,t)&=&\frac{1}{\pi^{1/4}}\sqrt{\frac{{\cal N}_j}{w_j}}
\exp[-(x-x_{0j})^2
/(2w_j^2)] \nonumber\\
&\times&
 \exp[i\{C_j(x-x_{0j})+\phi_j\}  ]        ,       
\end{eqnarray}
where 
$w_j$ are the widths  of component $j$  of localized BEC centered at
$x_{0j}$,   $\phi_j$ is the corresponding phase, $C_j$ is the linear phase 
coefficient and ${\cal N}_j$ is the normalization of component $j$. We shall 
consider solution where the widths are comparable to the OL wavelength. 
Generally, the variational approach is valid 
when the OL is only smooth and slowly varying on the localized states scale
\cite{bishop}. 
We shall obtain 
useful relations among the variational parameters of the 
localized BEC states. This will allow us to get a better physical 
understanding of the localized states.

The GP equations (\ref{eq1}) and (\ref{eq2}) can be derived from the 
following Lagrangian \begin{eqnarray}\label{lag} 
L(t)&=&\int_{-\infty}^\infty \biggr[\sum_{j=1}^2\biggr( \frac{i}{2}[ 
u_j^*\dot u_{j}- u_j\dot u^*_{j}]-\frac{1}{2}[|u'_{j}|^2+g_j|u_j|^4] 
\nonumber \\ &-&V(x)|u_j|^2\biggr) -g_{12}|u_1|^2|u_2|^2\biggr] dx, 
\end{eqnarray} 
where the overhead dot denotes time derivative, the star denotes 
complex conjugation,  and the 
prime denotes space derivative. Using  ansatz (\ref{va}) in Eq. 
(\ref{lag}) we get 
\begin{eqnarray}\label{lag2} L(t)&=&\sum_{j=1}^2{\cal 
N}_j\biggr[ (C_j\dot x_{0j}-\dot \phi_j)-\frac{1}{2} 
\left(\frac{1}{2w_j^2}+C_j^2\right)\nonumber \\ &-&\frac{g_j{\cal 
N}_j}{2\sqrt{2\pi}w_j} 
+\sum_{l=1}^2 L_{lj} \biggr]+L_{12},\nonumber \\ \\ 
L_{12}&\equiv &-\int_{-\infty}^{\infty}g_{12}|u_1|^2|u_2|^2dx\nonumber \\ &=&- 
\frac{g_{12}{\cal N}_1{\cal N}_2}{\sqrt{\pi(w_1^2+w_2^2)}}\exp\left[ 
-\frac{\rho^2}{w_1^2+w_2^2}\right],\label{l12} \\
L_{lj}&\equiv &\frac{A_l}{2}\left[ \cos (2k_l x_{0j})\exp(-k_l^2w_j^2)-1\right]
\label{llj}
\end{eqnarray} where 
$\rho=x_{02}-x_{01},$  and the variational 
parameters are the norm ${\cal N}_j$, width $w_j$, position $x_{0j}$, 
and phase parameters $C_j$ and $\phi_j$. We use  the Euler-Lagrange 
equation \begin{eqnarray} \frac{\partial L}{\partial 
\sigma}-\frac{d}{dt}\frac{\partial L}{\partial \dot \sigma}=0, 
\end{eqnarray} where $\sigma$ is the variational parameter. The first 
variational equation using $\sigma=\phi_j$ yields ${\cal N}_j =$ constant. 
Without losing generality we shall take this constant to be unity and 
use it in the subsequent equations. Taking $\sigma = x_{0j}, C_j, w_j,$ 
and ${\cal N}_j,$ respectively, we obtain the following equations 
\begin{eqnarray}\label{var1} \dot C_j &=& \frac{\partial}{\partial 
x_{0j}}\left[\sum_{l=1}^2 L_{lj}
+L_{12} \right], 
\\ 
\dot 
x_{0j}&=&C_j, \label{var2}  \\ \label{var3} 
\frac{1}{w_j}&+&\frac{g_j}{\sqrt{2\pi}}+\frac{2g_{12}w_j^3}{\sqrt 
\pi(w_1^2 +w_2^2)^{3/2}}\left[1-\frac{2\rho^2}{w_1^2+w_2^2} 
\right]\nonumber \\ &\times&\exp \left[-\frac{\rho^2}{w_1^2+w_2^2} 
\right]-\sum_{l=1}^2 2A_lw_j^3k_l^2 \cos(2k_lx_{0j})\nonumber \\ 
&\times&\exp(-k_l^2w_j^2)=0,\\ \label{var4} \dot \phi_j 
&=&\frac{C_j^2}{2}-\frac{1}{4w_j^2}-\frac{g_j}{w_j\sqrt{2\pi}} 
+\frac{\partial L_{12}}{\partial {\cal N}_j}\biggr|_{{\cal N}_j=1} 
+\sum_{l=1}^2 L_{lj}. \nonumber \\  \end{eqnarray} 
The set of equations (\ref{var1}) $-$ (\ref{var4}) shows 
that the phase $\phi_j$ has no effect on the location $x_{0j}$, width 
$w_j$ and the coefficient $C_j$ of solitons, whereas it is determined by 
the variational equation (\ref{var4}). Hence we shall neglect Eq.  
(\ref{var4}) in the following and consider only Eqs.  (\ref{var1}) $-$ 
(\ref{var3}). (However, we could not set phase $\phi_j=0$ in the beginning, as 
 the Euler-Lagrange equation for  $\phi_j$ leads to the important result 
of norm conservation.)
Equation (\ref{var3}) determines the width $w_j$ in terms 
of nonlinearities $g_j$, $g_{12}$ and the parameters of the 
bi-chromatic OL potential. 
The density distribution is quasi-Gaussian only for small values of $x_{0j}$,
and we shall be limited to this constraint (small $|x_{0j}|$) 
while considering the 
variational approach. In the symmetric case, while $g_1=g_2$, the widths 
of the two localized states are equal. However, when $g_1\ne g_2$, the 
widths of the two localized states are not equal and one has two asymmetric 
localized states. 
When $x_{01}=x_{02}=0$, the widths of the two stationary BEC localized states 
are determined by 
\begin{eqnarray}\label{varw}
\frac{1}{w_j}&+&\frac{g_j}{\sqrt{2\pi}}+\frac{2g_{12}w_j^3}{\sqrt
\pi(w_1^2 +w_2^2)^{3/2}}\nonumber \\  
&-&\sum_{l=1}^2 2A_lw_j^3k_l^2 
\exp(-k_l^2w_j^2)=0.
\end{eqnarray}

Now we obtain, following Ref. \cite{cheng}, 
 the equation of motion describing the center of the two 
BEC localized states as if they were a particle in an effective 
potential. Thus, we can determine the motion of equivalent particles 
using the classical mechanics analogy. Inserting Eqs. (\ref{l12}), (\ref{llj}) and  
(\ref{var1}) into Eq. (\ref{var2}), we obtain the following equation
for dynamics 
\begin{eqnarray}\label{effpot2}
\frac{d^2x_{0j}}{dt^2}&=&-\frac{\partial V_{\mathrm{eff}j}}{\partial x_{0j} }
\; , \\
V_{\mathrm{eff}j}&=& \frac{g_{12}}{\sqrt{\pi(w_1^2+w_2^2) }}
\exp\left[-\frac{\rho^2}{w_1^2+w_2^2}  \right]\nonumber \\
&-&  \sum_{l=1}^2 \frac{A_l}{2}[\cos(2k_lx_{0j})
\exp(-k_l^2w_j^2)-1]\; ,\label{effpot}
\end{eqnarray}
 where the an-harmonic effective potentials $V_{\mathrm{eff}j}$ are the 
dynamic potentials for the movement of the center of the localized BECs 
and depend on characteristics of the two BEC localized states (e.g., 
width $w_j$ and position $x_{0j}$) and potential parameters
(e.g., $k_l$ and $A_l$). The effective potentials 
have two parts. The first term on the right of Eq. (\ref{effpot}) is 
induced by the mutual interaction of the two BEC localized states. When 
the coupling constant $g_{12}>0$, it is positive and contributes 
to a  
{{repulsive potential barrier 
for   BEC localized states.}} The second term in Eq. (\ref{effpot})
arises from the bi-chromatic OL potential and contributes
to an attractive potential well,  if $|x_{0j}|$ is small enough. 
The combination of the two terms may lead to local minima or local 
maxima at the origin $x_{0j}=0$
in the effective potential $V_{\mathrm{eff}j}$. 

\begin{figure}
\begin{center}
\includegraphics[width=.49\linewidth]{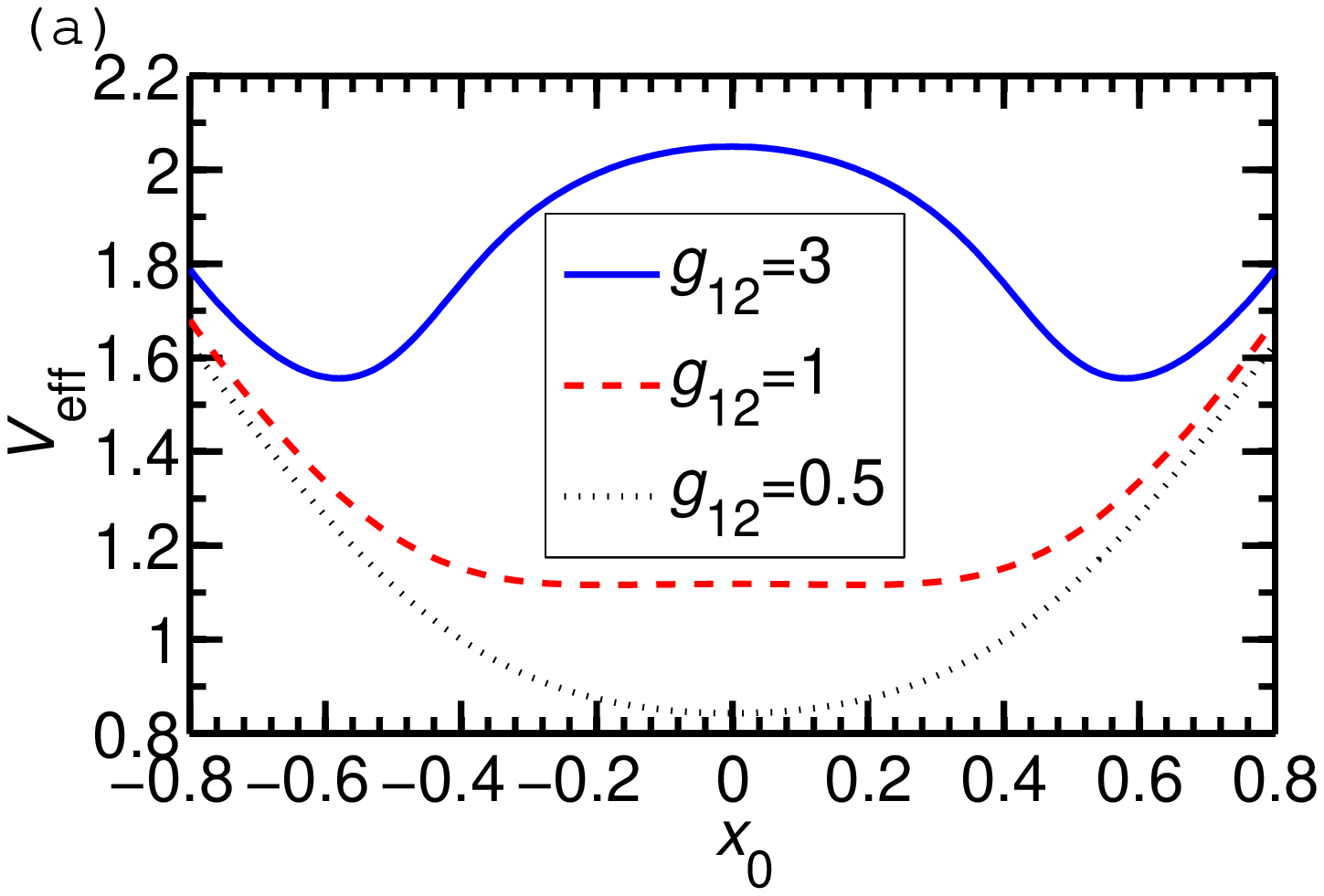}
\includegraphics[width=.49\linewidth]{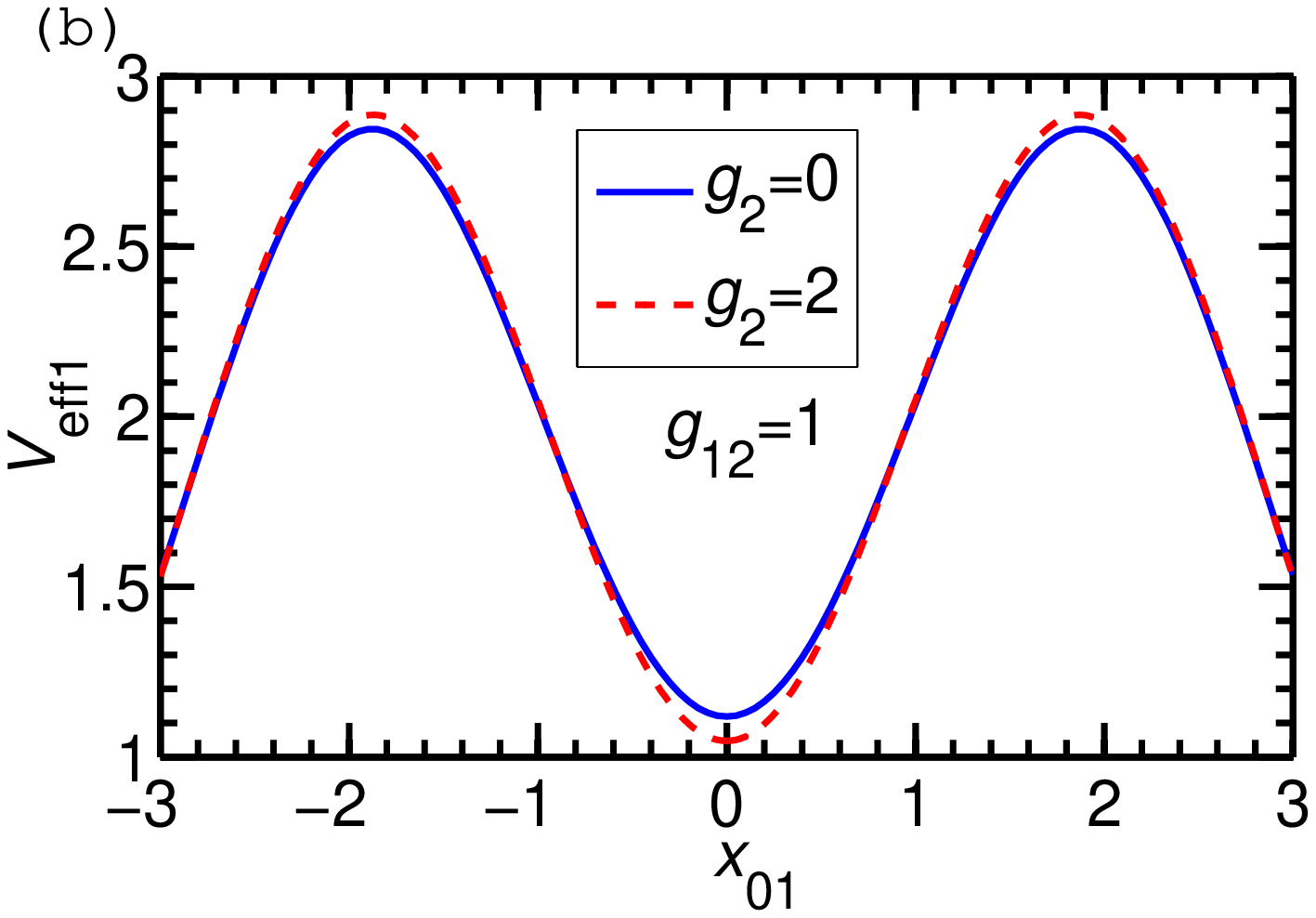}
(a)  \hskip 3 cm   (b)
\end{center}

\caption{(Color online) (a) The effective potential $V_{\mathrm{eff}}(x_0)$ vs. 
the position of the localized state $x_0\equiv x_{01} = -  x_{02}$ 
in the symmetric case $g_1=g_2
=0$ for $g_{12}=0.5,1$ and 3 as obtained from Eq. (\ref{effpot}).  
The respective widths are calculated from Eq. (\ref{var3}). (b) 
The effective potential felt by the first component 
$V_{\mathrm{eff}1}(x_{01})$ vs. its position $x_{01}$ as obtained from 
Eq.   (\ref{effpot}). The widths of the two states are calculated from 
Eq. (\ref{varw}).
In both cases $-$ (a) and (b) $-$ the potential parameters are 
 $\lambda_1=8, \lambda_2=0.862\lambda_1, s_1=8, s_2=2.4.$
}
\label{figpot}
\end{figure}

When the two BEC localized states are symmetrical $(g_1=g_2)$, they will 
feel the same 
effective potential. In that case,  we should have 
$w_1=w_2=w$ and $x_{01}=-x_{02}=x_0$. 
The effective potential $V_{\mathrm{eff}}$
is plotted in Fig. \ref{figpot} (a) vs. $x_0$ 
in the symmetric case  $g_1=g_2=0$
for different inter-species interaction: $g_{12}=0.5,1,$ and 3.
Equation (\ref{var3}) is first solved to find $w$ as a function of $x_0$. 
This result is then substituted in Eq. (\ref{effpot}) to find 
$V_{\mathrm{eff}}(x_0)$  as plotted in Fig. \ref{figpot} (a).
From  Fig. \ref{figpot} (a) we find that
if $g_{12}$ is smaller, the effective potential possesses a local minimum 
at $x_0=0$. In this case if
we start with two BEC localized states at $x=0$ for small 
$g_{12}$, they 
will be in stable equilibrium and unsplit. The effective 
potential becomes weaker with the local minimum
at $x_0=0$ less pronounced 
as $g_{12}$ increases.  If $g_{12}$ is 
large enough (e.g., $g_{12}=3$ in Fig. \ref{figpot} (a)), 
a local maximum appears in the effective potential
at $x_0=0$; and the peak value at maximum 
increases as $g_{12}$ increases. 
In this case  two  overlapping (unsplit) initial localized  BEC states at  
the maximum at $x=0$  will be in a  metastable configuration. 
When they are slightly perturbed, for example, 
by slightly moving the BEC center(s) from $x=0$, 
they will move respectively toward the minima of the effective potential 
next to the 
maximum and split into two non-overlapping BECs localized at the two local 
minima of the effective potential 
on both sides of $x_0=0$ as we shall see in Sec. \ref{IIIA}.

When the two BEC localized states are asymmetrical ($g_1\ne g_2$), they 
will feel different effective potentials. It is difficult to calculate 
these effective potentials exactly. But we can calculate the effective 
potential(s)  (\ref{effpot}) under simplifying assumptions. The widths 
$w_1$ and $w_2$ are assumed to be independent of positions $x_{01}$ and
 $x_{02}$ and were calculated using Eq. (\ref{varw}) and substituted 
into Eq. (\ref{effpot}). The function $V_{\mathrm{eff}1}(x_{01})$ so 
obtained is plotted in Fig. (\ref{figpot}) (b) vs. $x_{01}$, for $g_1=0, 
g_{12}=1, g_2=0$ and 2. The effective potential acting on component 1 is 
stronger as $g_2$ is increased from 0 to 2. This will lead to a reduction 
of the width of the corresponding localized state as $g_2$ 
is increased from 0 to 2 as we shall see in Sec. \ref{IIIA}.

\section{Numerical Results}

\label{III}

We perform {the} numerical simulation employing the real-time split-step 
Fourier spectral method  with space step 0.04, 
time step 0.001.  (We also checked the results using the real-time
Crank-Nicolson discretization routine, 
the FORTRAN programs  which are given in Ref. \cite{bo1}.) Because of the 
oscillating nature of the potential, great care was needed to obtain a 
precise { localized state}. The accuracy of the numerical simulation was 
tested by varying the space and time steps as well as the total number 
of space steps. 
The initial input pulse is a Gaussian wave packet, 
$u_{j}(x,t=0)=\pi^{-1/4}\exp (-x^2/2)$
 with a parabolic trap $V'(x)=x^2/2$. Let 
$g_1=g_2=g_{12}=0$ and the coupled Eqs.  (\ref{eq1}) and (\ref{eq2})  
become the linear 
Schr\"odinger 
equation. To start the numerical simulation
  the parabolic trap is slowly turned off and 
the bi-chromatic OL is slowly turned on; 
the increment in 
the coefficient $s_i$ in each time step  
is 
0.00005.  Successively, we add gradually the 
nonlinear coefficient $g_2$ and the coupling constant $g_{12}$  slowly. 
 The increment of $g_2$ and $g_{12}$ is 0.00001 in each time step. 
Thereby the stationary 
localized states are obtained.

To perform a systematic numerical study of localization
we take $s_1=8$, $s_2=2.4$,
$\lambda_1 =8$ and $\lambda_2=0.862\lambda_1$ 
maintaining 
the ratio  $\lambda_2/\lambda _1 =0.862$ 
similar to the value employed in the experiment 
\cite{roati}. 
The values of $s_1$ and $s_2$ are also in the range 
employed in the experiment. With these set of parameters and for 
weak inter-species and intra-species interactions,  
the localized 
BEC state could be  mostly confined in the single cell of the bi-chromatic 
OL potential. 
All results reported here are obtained with these 
potential parameters except those in Figs. \ref{prob1} (e) and (f) 
where we study the effect of the change of potential parameters.
We consider a transverse harmonic oscillator length 
$a_\perp \approx 1 $ $\mu$m, so that the
 wave lengths of the OL potential are approximately $ 800$ nm and $ 689$ 
nm. We shall take $g_1=0$ throughout the present investigation.

\subsection{Localized states with potential (\ref{pot1}) 
on both components}

\label{IIIA}

\begin{figure}
\begin{center}
\includegraphics[width=.49\linewidth]{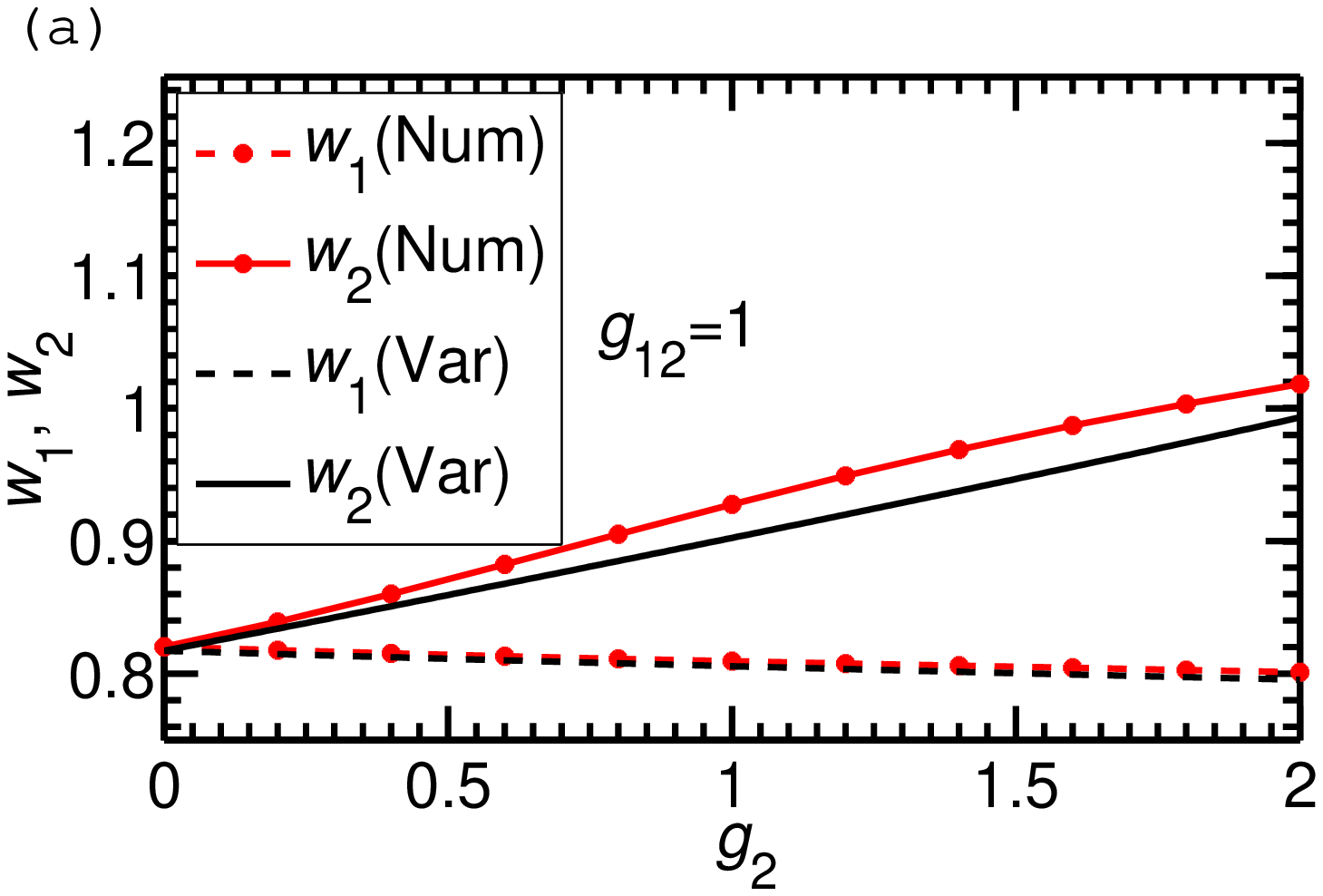}
\includegraphics[width=.49\linewidth]{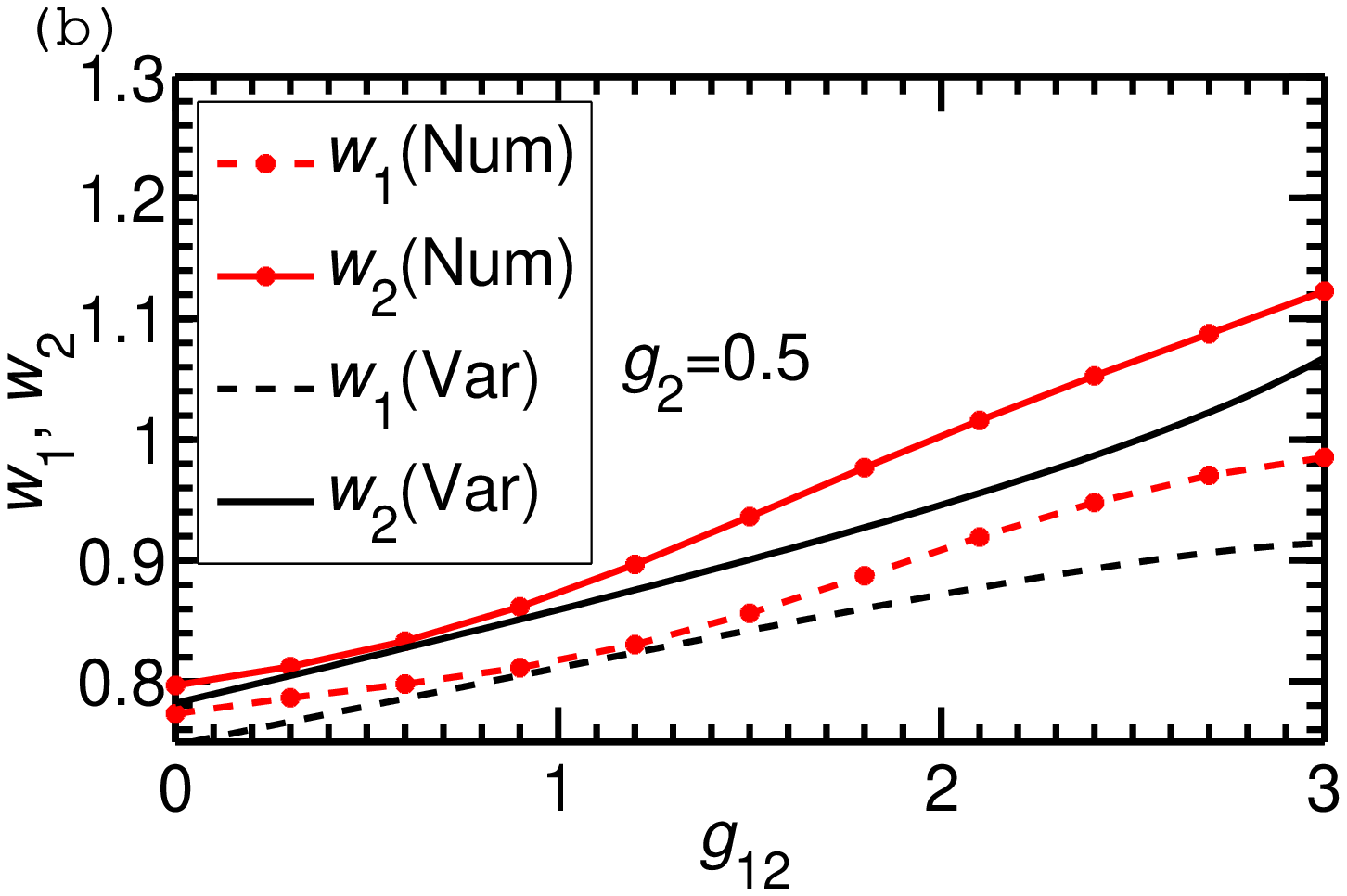}
(a)  \hskip 3 cm   (b)
\end{center}

\caption{(Color online) (a) The numerical and variational 
widths $w_1$ and $w_2$ of the two stationary BEC localized 
states vs. $g_2$ for $g_{12}=1,g_1=0$. (b) The same as in (a) vs. $g_{12}$ 
for $g_2=0.5,g_1=0$. The potential (\ref{pot1}) acts on the two components 
where the 
parameters are $\lambda_1 =8, \lambda_2 
=0.862\lambda_1, s_1=8,  s_2=2.4$. 
}
\label{width}
\end{figure}

We start the numerical analysis with a consideration of the widths of 
the two components of the localized BECs
with potential (\ref{pot1}) acting on both components. 
In this  case both the components could have a maximum at the center 
$x=0$ with localized states of quasi-Gaussian forms and when this 
happens the system can be described well by the variational approximation. 
Assuming a Gaussian distribution for the localized BEC, the numerical width
 is calculated via 
\begin{eqnarray}
w_j^2=2\int_{-\infty}^\infty |u_j|^2 x^2 dx.
\end{eqnarray}
The numerical and variational results for the widths $w_j$ 
vs. $g_2$ for 
$g_{12}=1$ are plotted in Fig. \ref{width} (a). The same for $g_2=0.5$
vs. $g_{12}$ are plotted in Fig. \ref{width} (b).   
In these plots there is reasonable agreement between the numerical and 
variational results for the width and the width usually 
increases with the increase of nonlinearities $g_2$ and $g_{12}$ except in 
the case of $w_1$ in Fig. \ref{width} (a). The general 
increase of width with the increase of non-linearity (increase of repulsion) 
is expected, however, the decrease of $w_1$ with $g_2$ in Fig. \ref{width} (a)
requires special 
attention.  This phenomenon will be clear if we consider the effective
potential of component 1 as illustrated in Fig. \ref{figpot} (b) when 
$g_2$ is changed from 0 to 2. We find that the effective potential becomes 
slightly stronger as $g_2$ increases corresponding to a reduction 
of width $w_1$ in Fig. \ref{width} (a) with the increase of $g_2$.

\begin{figure}
\begin{center}
\includegraphics[width=.49\linewidth]{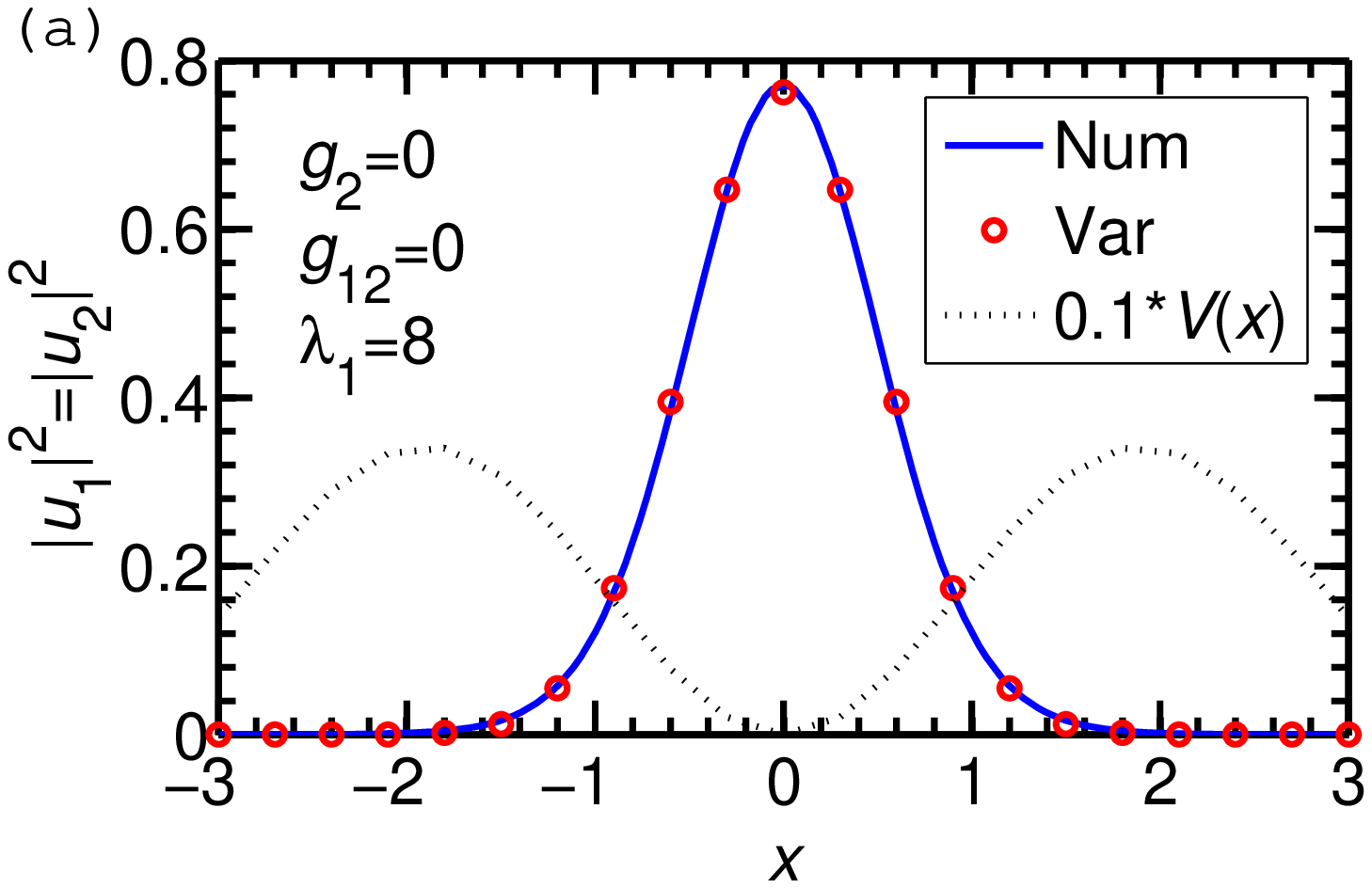}
\includegraphics[width=.49\linewidth]{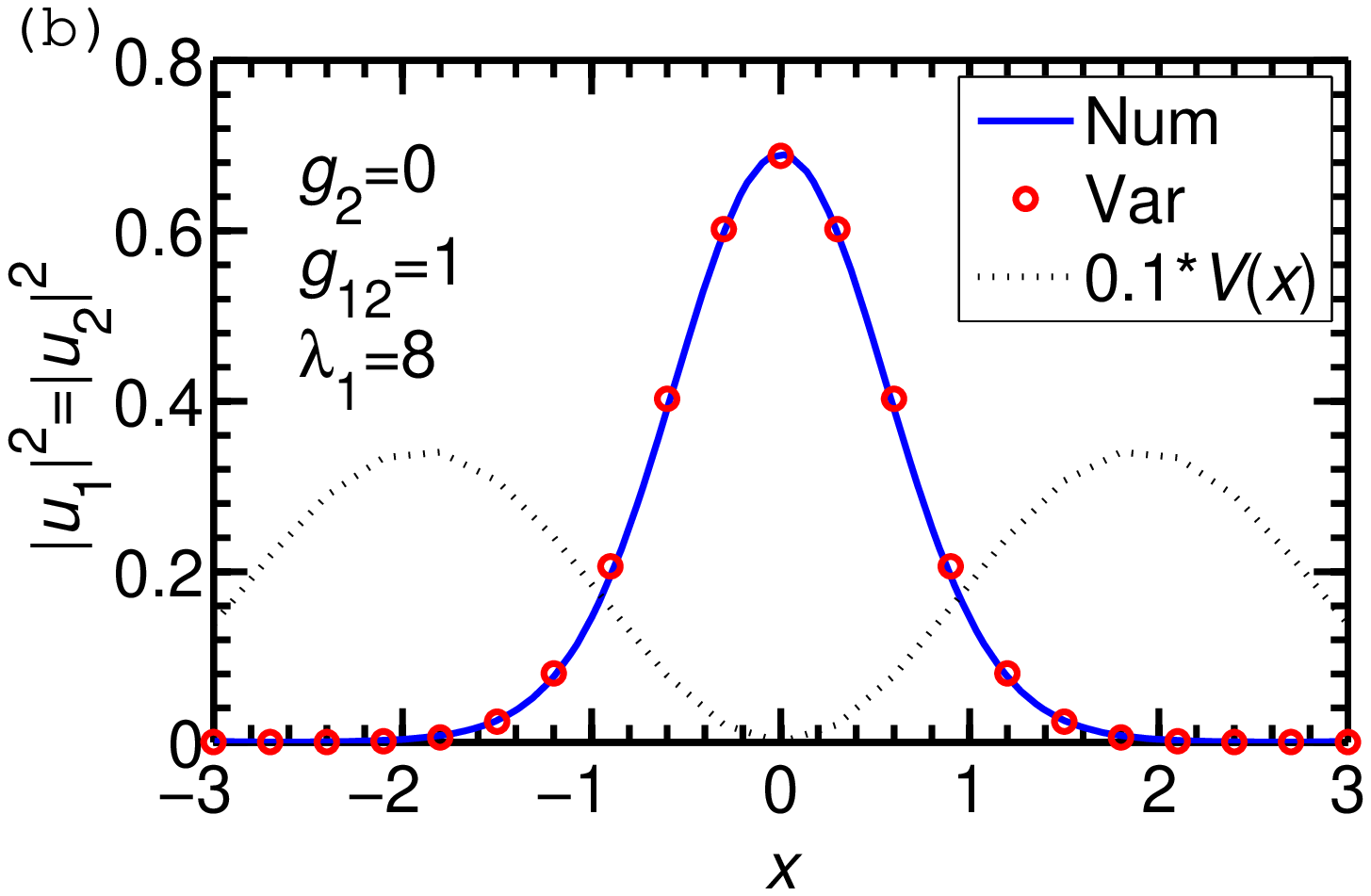}
(a)  \hskip 3 cm   (b)   \hskip 3 cm
\includegraphics[width=.49\linewidth]{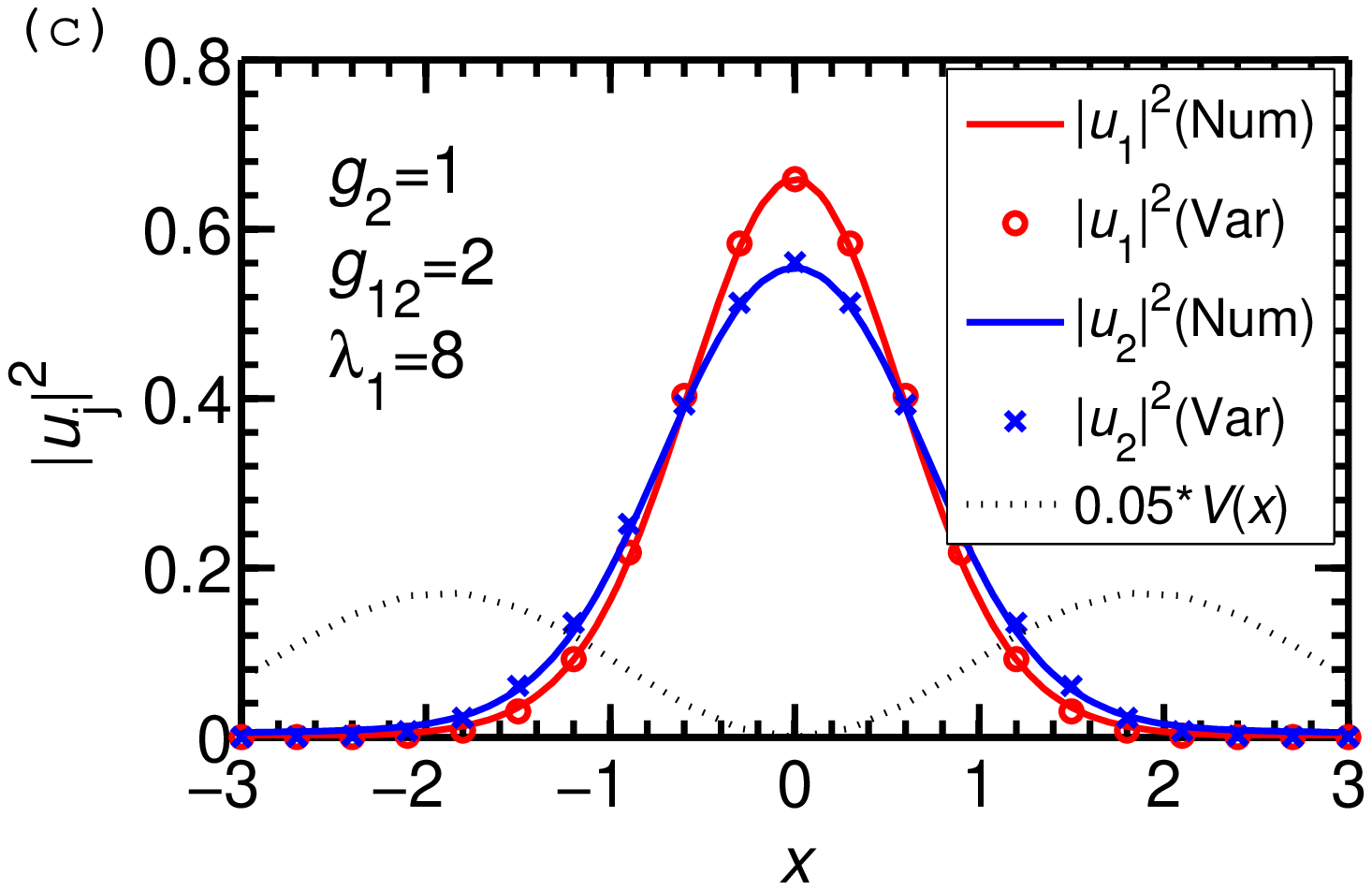}
\includegraphics[width=.49\linewidth]{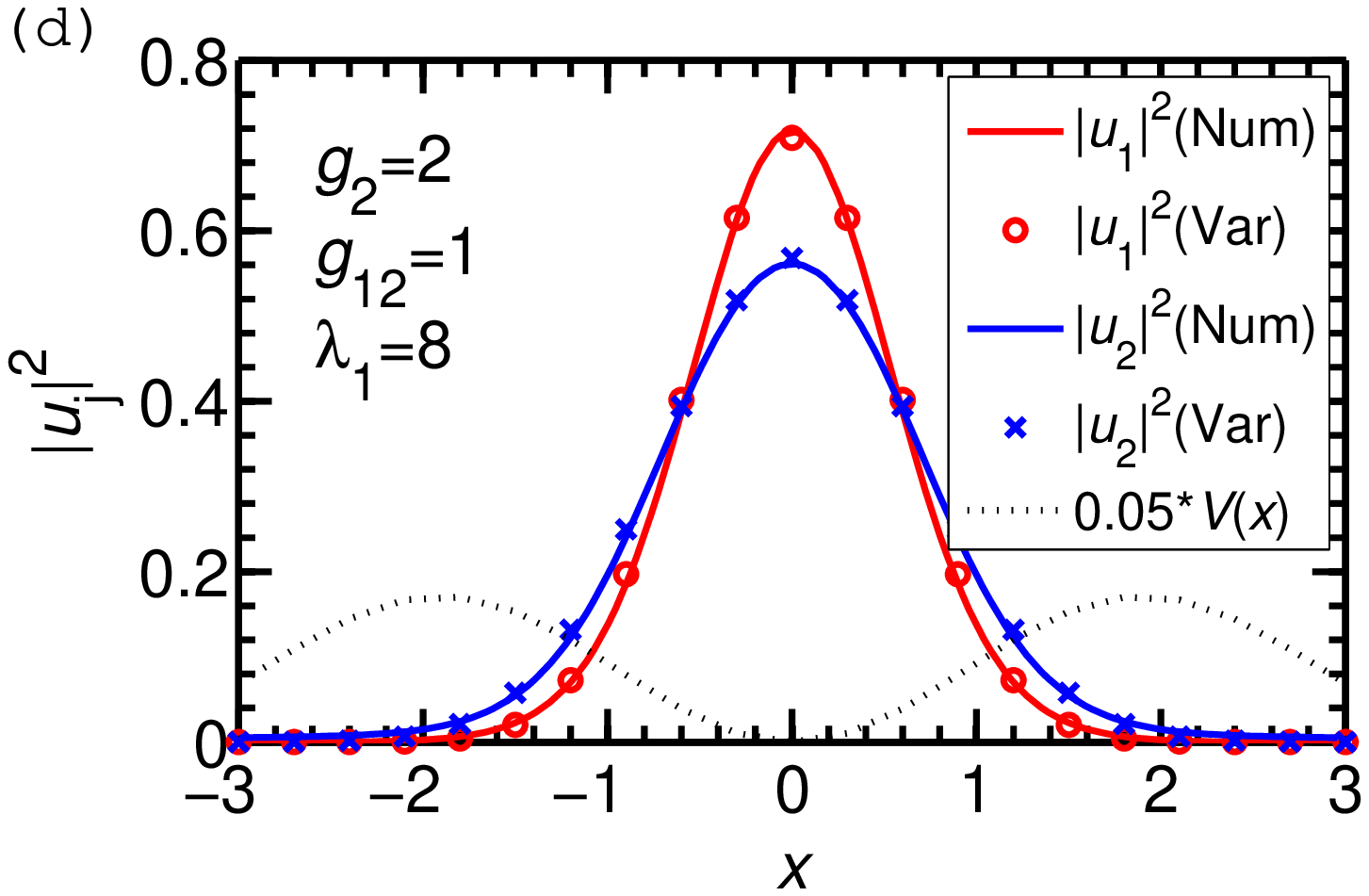}
(c)  \hskip 3 cm   (d)  \hskip 3 cm
\includegraphics[width=.49\linewidth]{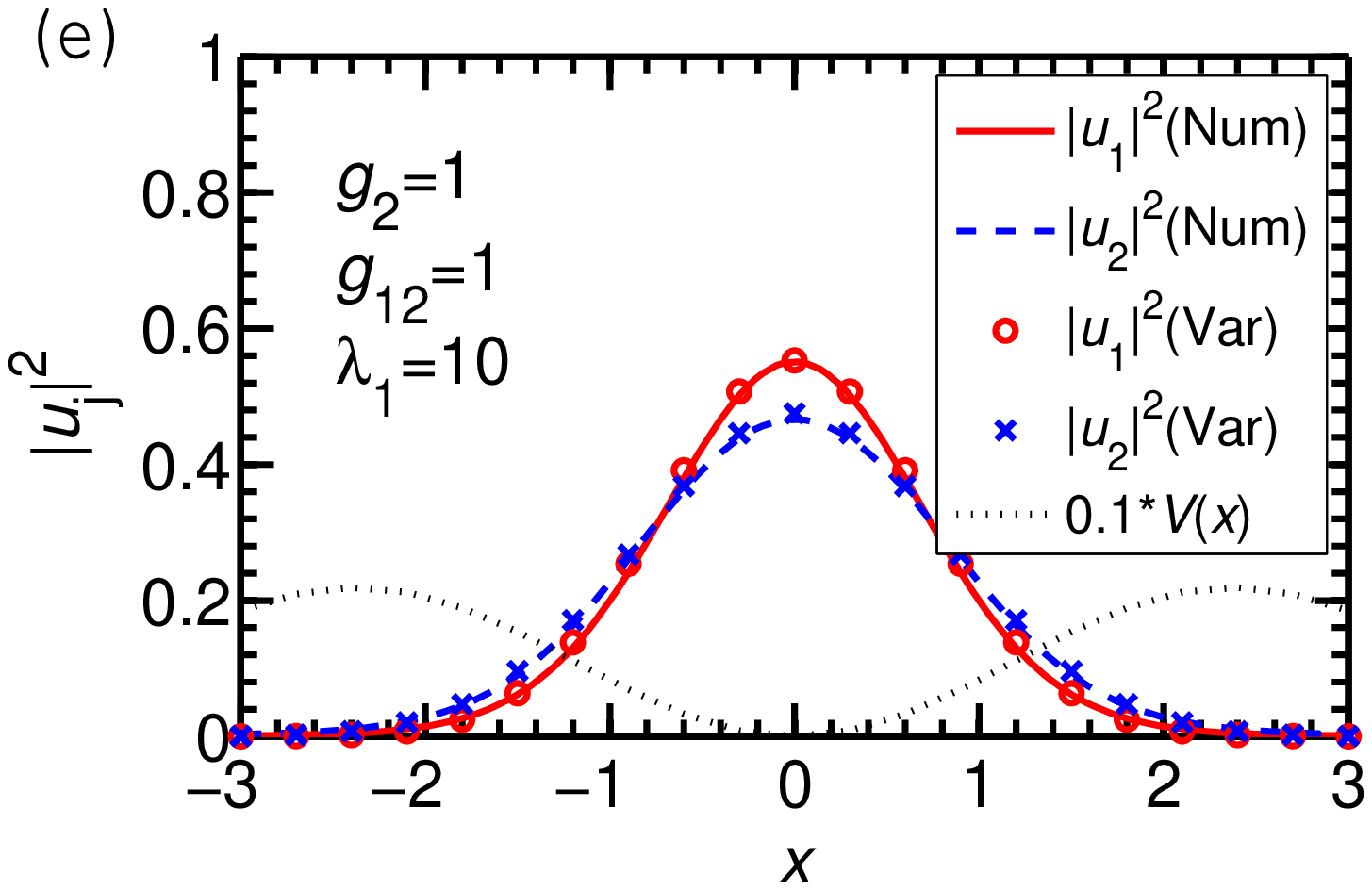}
\includegraphics[width=.49\linewidth]{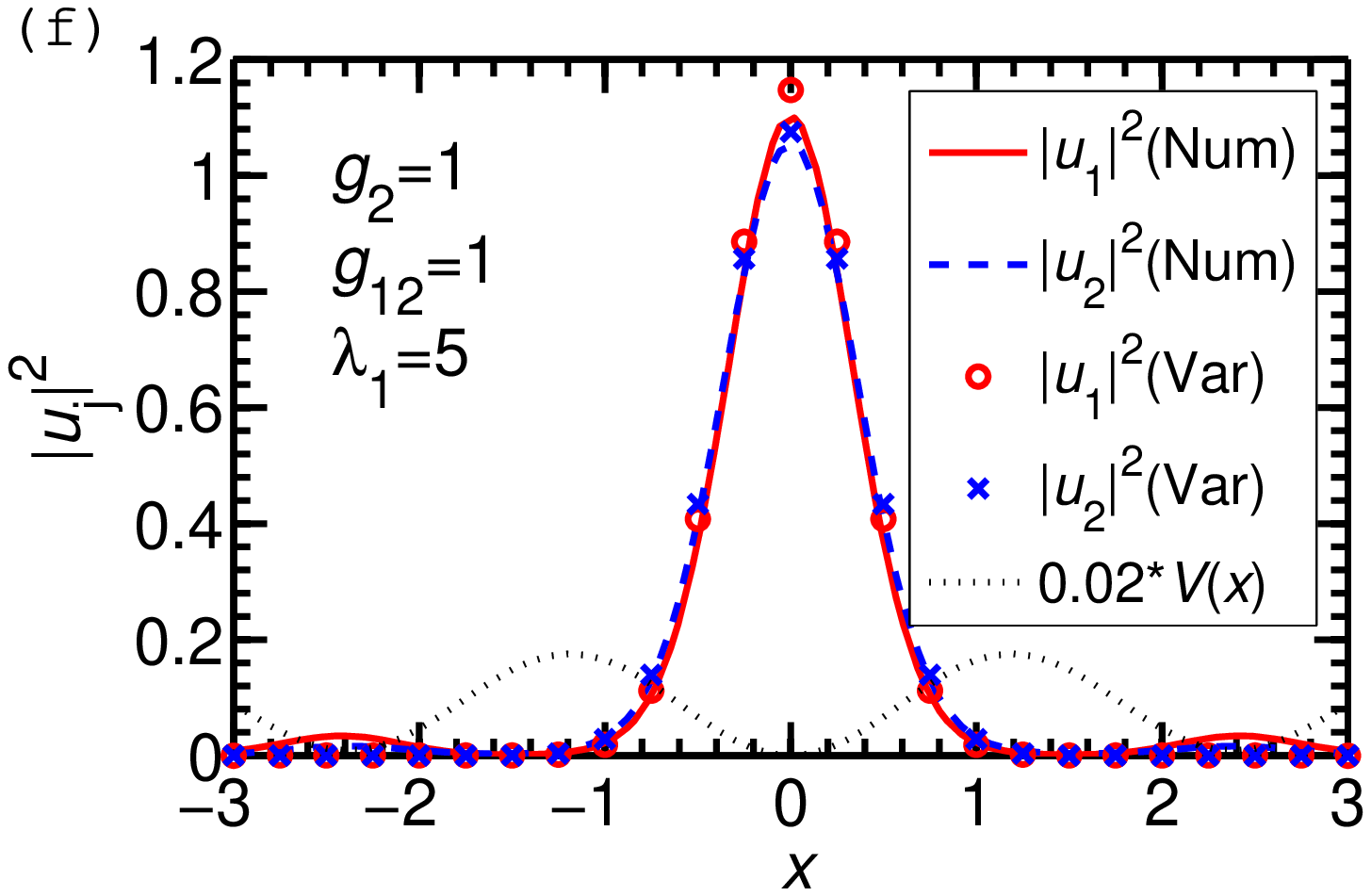}
(e)  \hskip 3 cm   (f)
\end{center}

\caption{(Color online) The densities $|u_1(x)|^2$ and $|u_2(x)|^2$
of the two localized BEC states vs. $x$  for  (a) 
$g_1=g_2=g_{12}=0$;   
(b) $g_1=g_2=0$ and $g_{12}=1$; (c) $g_1=0,g_2=1,$ and $g_{12}=2$;
(d) $g_1=0, g_2=2,$ and  $g_{12}=1$. In (a), (b), (c), and (d) potential 
(\ref{pot1}) acts on the two components with  
parameters  $\lambda_1=8, \lambda_2=0.862\lambda_1, s_1=8, s_2=2.4.$
 The same as in (a)  for  $g_1=0, g_2=g_{12}=1$ with potential parameters
(e)
$\lambda_1=10, \lambda_2=0.862\lambda_1, s_1=8, s_2=2.4  $
and
(f)
$\lambda_1=5, \lambda_2=0.862\lambda_1, s_1=8, s_2=2.4.$
In each case the profile of the bi-chromatic OL potential is also shown.
}

\label{prob1}
\end{figure}

Next we consider typical  numerical 
profiles of the density of the localized BEC 
states $|u_j|^2$
and compare them with variational results for small values of $g_{12}$
when the centers of both components are at the origin: $x_{01}=x_{02}=0$.
In Figs. \ref{prob1} (a) and (b) we present the results for $g_1=g_2=0$ 
and $g_{12}=0$ and 1, respectively,  
while the two density profiles $|u_1|^2$ and $|u_2|^2$ are equal. 
Next we consider 
results for the asymmetric case 
while  $g_1\ne g_2$ while the two densities  are different.
In Figs. \ref{prob1} (c) and (d) we present the results for $g_1=0, g_2=1, 
g_{12}=2$ and for $g_1=0, g_2=2, g_{12}=1$, respectively. In both cases the 
component 1 is more strongly localized 
  in space with a smaller width. Next we consider the effect of a 
variation of  the wavelengths $\lambda_1$ and $\lambda_2$ on the 
densities. In Fig.  \ref{prob1} (e) we plot the densities  for 
$\lambda_1=10, \lambda_2=0.862 \lambda_1$, and in  Fig.  \ref{prob1} (f)
we plot the densities for
$\lambda_1=5, \lambda_2=0.862 \lambda_1$; in both cases we take 
$g_1=0,g_2=g_{12}=1, s_1=8,s_2=2.4.$
In Fig.  \ref{prob1} (e) with the increase of $\lambda$'s the central OL 
site has a larger spatial extension, consequently the 
densities have a 
smaller value at the maxima at the origin corresponding to a larger 
width. In Fig.  \ref{prob1} (f) 
with the decrease of $\lambda$'s the central OL 
site has a smaller spatial extension, consequently the 
densities have a
larger value at the maxima at the origin corresponding to a 
smaller width.
The plots in Figs. \ref{prob1} (a) $-$ (e)
indicate that the numerical results for densities  
are in perfect agreement with 
the variational results. It also indicates  the good accuracy of the 
numerical routine. In these cases the density has a quasi-Gaussian profile 
used in the variational analysis. In Fig. \ref{prob1} (f) the numerical 
densities have secondary maxima on both sides of the central peak, and 
the profile deviates from the Gaussian and the agreement with the variational
results in this case is only fair.

\begin{figure}
\begin{center}
\includegraphics[width=.49\linewidth]{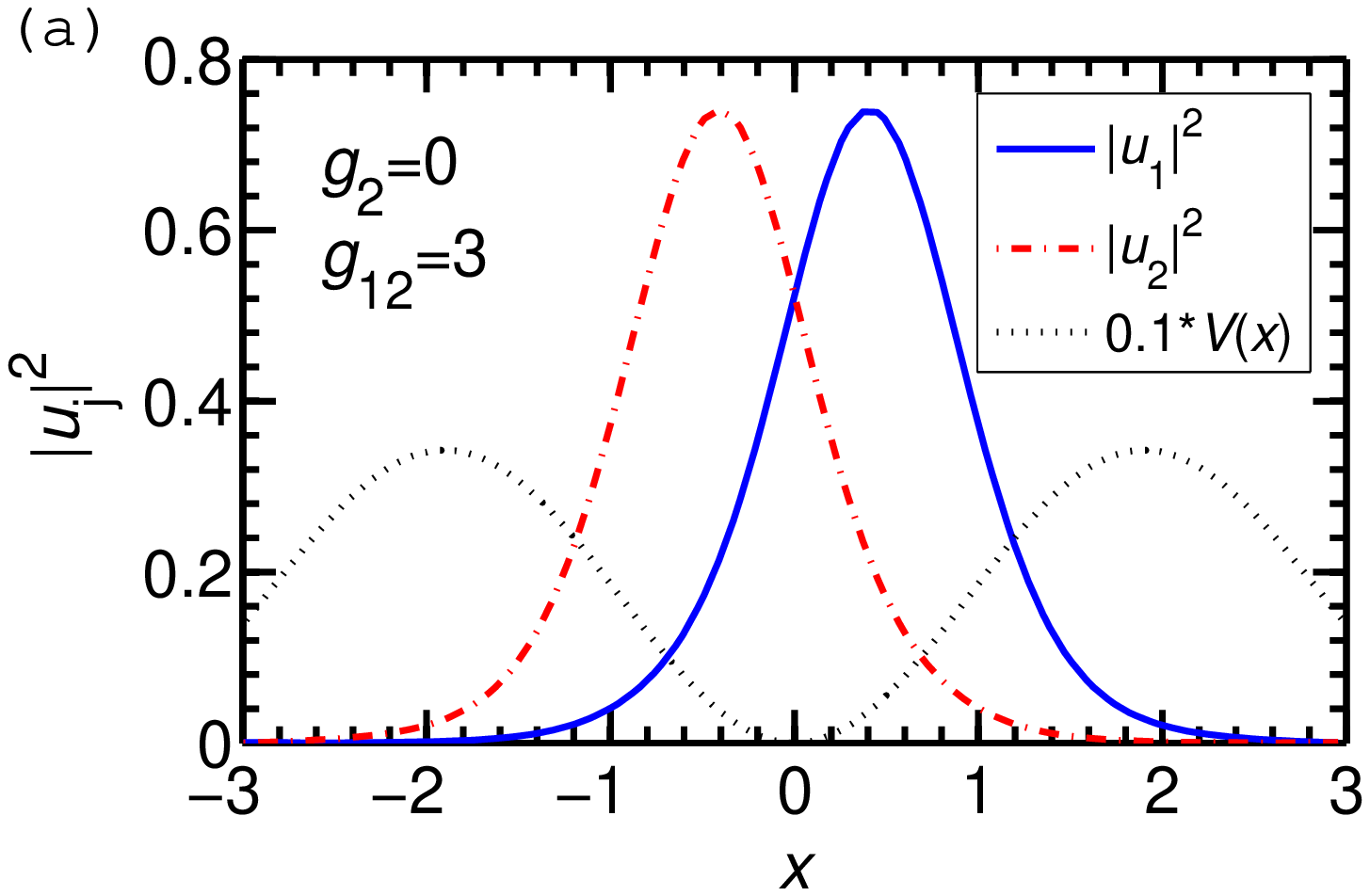}
\includegraphics[width=.49\linewidth]{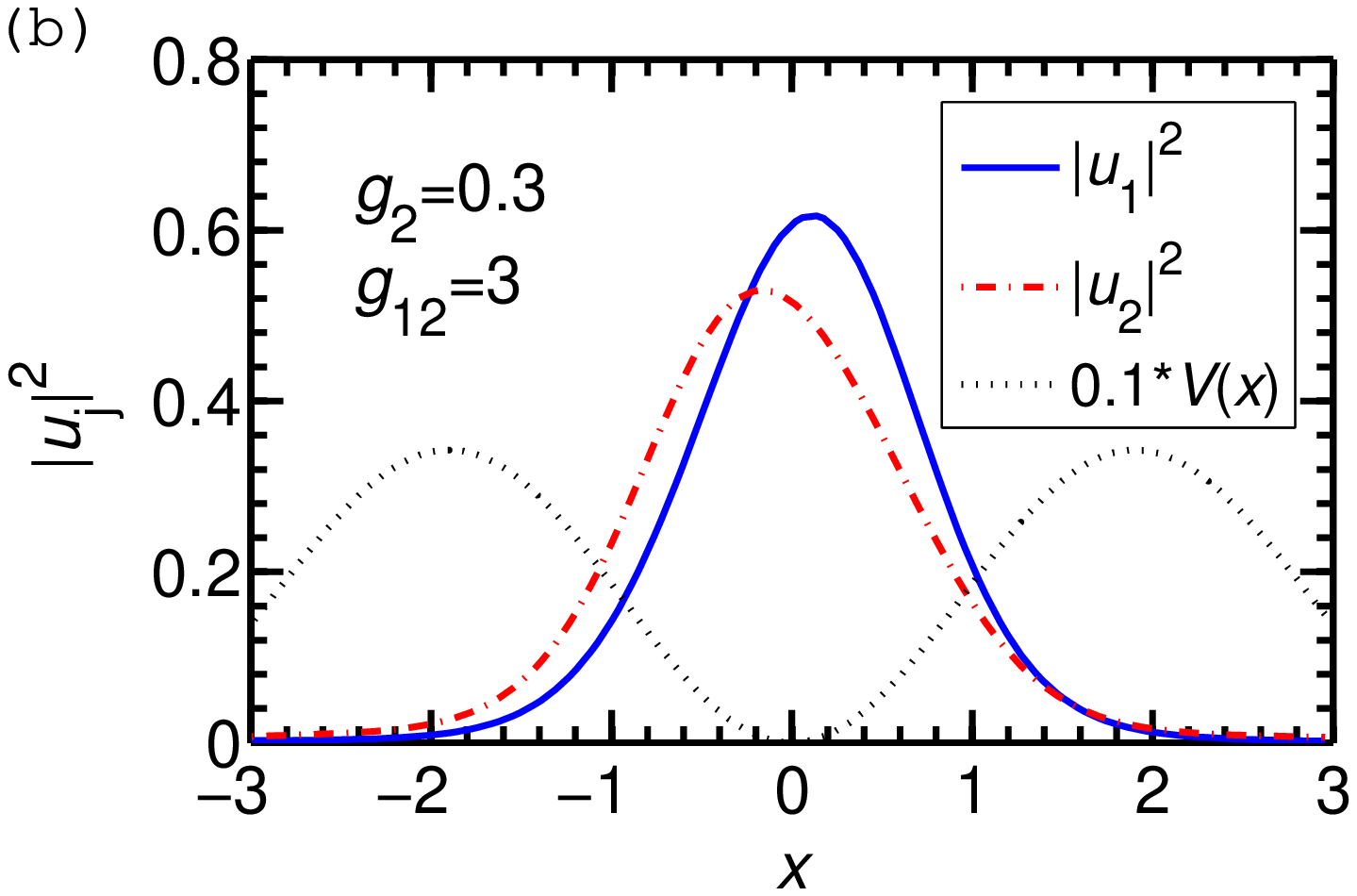}
(a)  \hskip 3 cm   (b)
\end{center}

\caption{(Color online) (a) The same as in Fig. \ref{prob1} (a) 
for (a) $g_1=g_2=0$ and $g_{12}=3$, and (b) $g_1=0,g_2=0.3$ and $g_{12}=3.$
The parameters of the  
potential are  $\lambda_1=8, 
\lambda_2=0.862\lambda_1, s_1=8, s_2=2.4$. 
}
\label{prob2}
\end{figure}

So far we presented results in the asymmetric case for small values of 
$g_{12}$ where the two localized states are overlapping (unsplit) and 
centered at $x=0$. As $g_{12}$ increases it is clear from Fig. 
\ref{figpot} (a) that the effective potentials experienced by the 
components develop a maximum at the origin. Hence the position $x=0$ 
ceases to be one of stable equilibrium for the components. Consequently, 
the two localized BECs move on opposite sides of the point $x=0$ to 
attain split (separated) configurations of stable equilibrium.  
We illustrate this 
in the symmetric case in Fig. \ref{prob2} (a) for $g_1=g_2=0, g_{12}=3$ 
where we plot the densities of the two split components. The two 
densities continue to be symmetric: $|u_1(x)|^2=|u_2(-x)|^2$ 
with centers at 
$|x_{01}|=|x_{02}|=0.45$. From Fig. \ref{figpot} (a) we find that the 
minima of the effective potential for the same set of parameters as in 
Fig. \ref{prob2} (a) are at $|x_{01}|=|x_{02}|=0.58$ indicating the 
centers of the split solitons to be at $\pm 0.58$. The variational analysis 
is valid for small splitting, and considering that 
the value $\pm 0.58$ indicating splitting is not too small 
compared to the extension of the localized states, the agreement between 
the numerical displacement 0.45 of the localized state and its variational 
estimate of 0.58  should be considered satisfactory.  In Fig. 
\ref{prob2} (b), 
we present an example of the localized asymmetric split states for 
$g_1=0, g_2=0.3, g_{12}=3$, where the two states have different (asymmetric)
densities.

\begin{figure}
\begin{center}
\includegraphics[width=\linewidth]{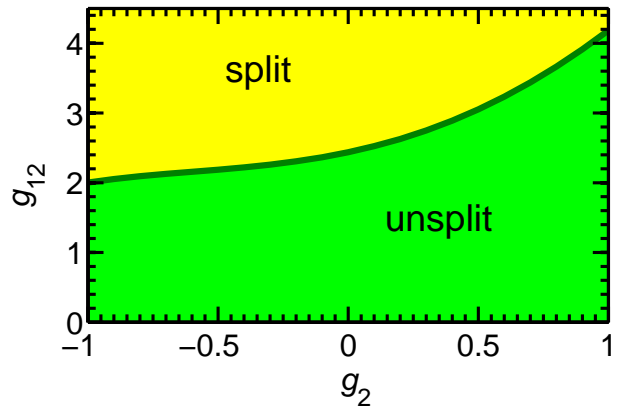}
\end{center}

\caption{(Color online) The phase diagram of  $g_{12}$ vs. $g_2$  for $g_1=0$
showing the split and unsplit configurations of the two localized states
while potential (\ref{pot1}) is
effective on both components. 
The parameters of the  
potential are  $\lambda_1=8, 
\lambda_2=0.862\lambda_1, s_1=8, s_2=2.4$. 
}
\label{phase1}
\end{figure}

As the split and the unsplit configurations of the localized 
states are of interest, it is appropriate to represent the split and  
unsplit configurations in a $g_2$ vs. $g_{12}$ phase diagram 
 for $g_1=0$. This is illustrated in Fig. \ref{phase1}. 
{{(Similar split-unsplit phase diagram has also been studied 
in the context of symbiotic gap solitons \cite{malomed}.)}}
For small 
$g_{12}$, as expected, the localized states are in an overlapping (unsplit) 
configuration. The split configuration is achieved when $g_{12}$ is larger 
than a critical value indicating a minimum of inter-species
repulsion needed to move 
the localized states from $x=0$. Splitting is favored when the inter-species 
interaction  
is strongly repulsive with a large positive value of $g_{12}$ 
and when the intra-species interaction is attractive corresponding to a 
negative  $g_2$ as seen
in Fig. \ref{phase1}.

\subsection{Localized states with potential (\ref{pot1}) on component 1
and potential (\ref{pot2}) on component 2}

\label{IIIB}

\begin{figure}
\begin{center}
\includegraphics[width=.49\linewidth]{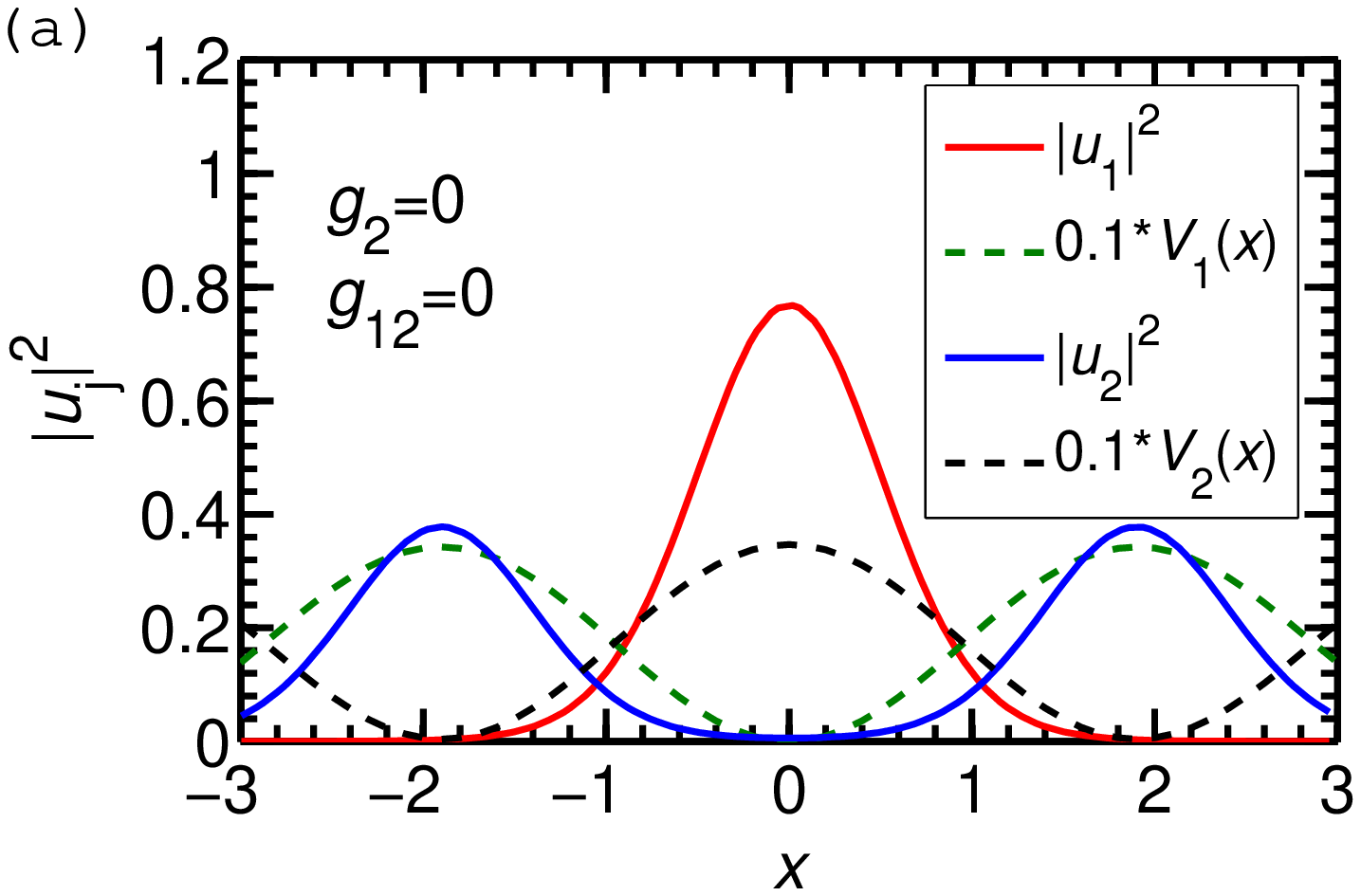}
\includegraphics[width=.49\linewidth]{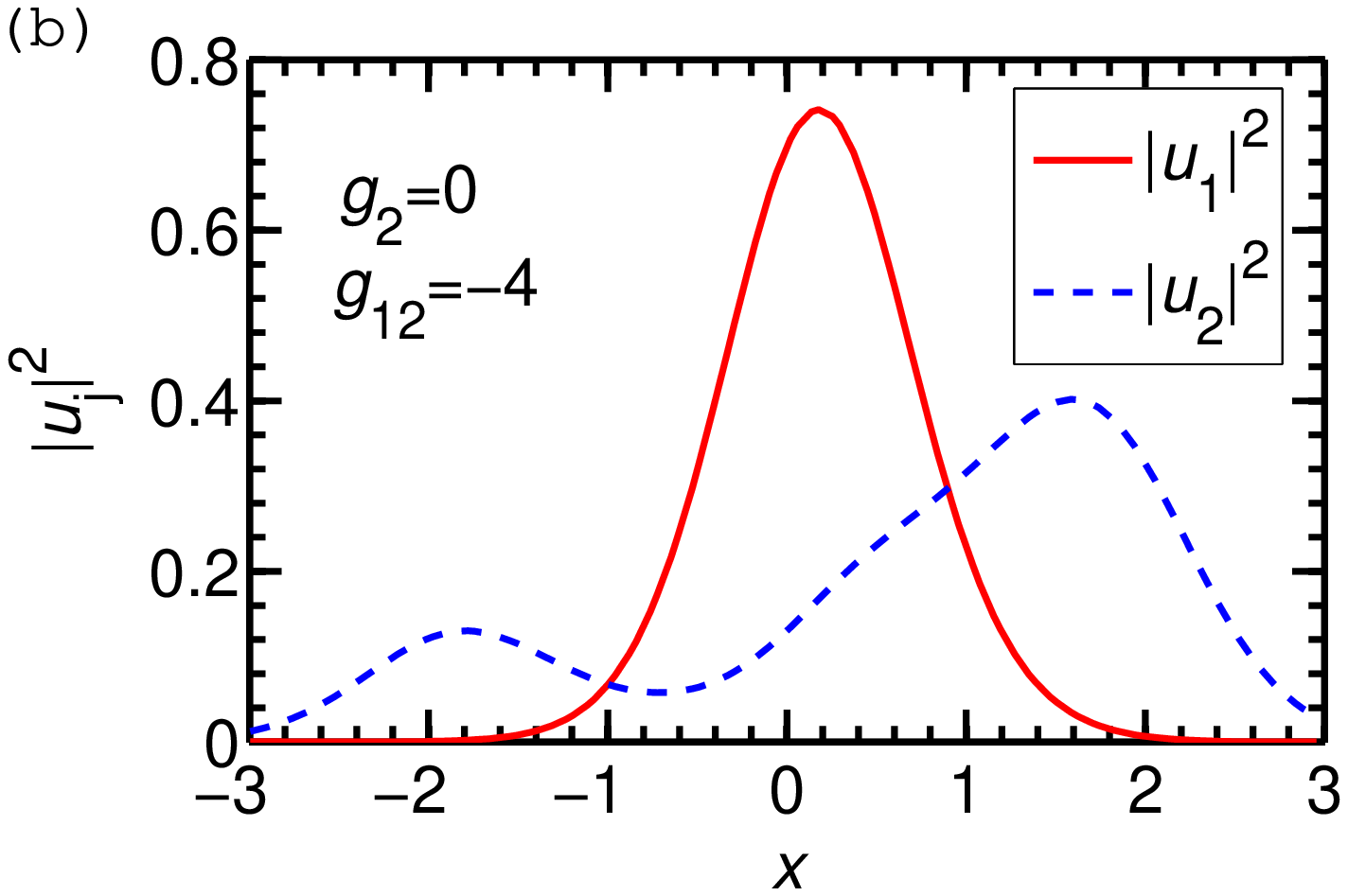}
(a)  \hskip 3 cm   (b) \hskip 3 cm 
\includegraphics[width=.49\linewidth]{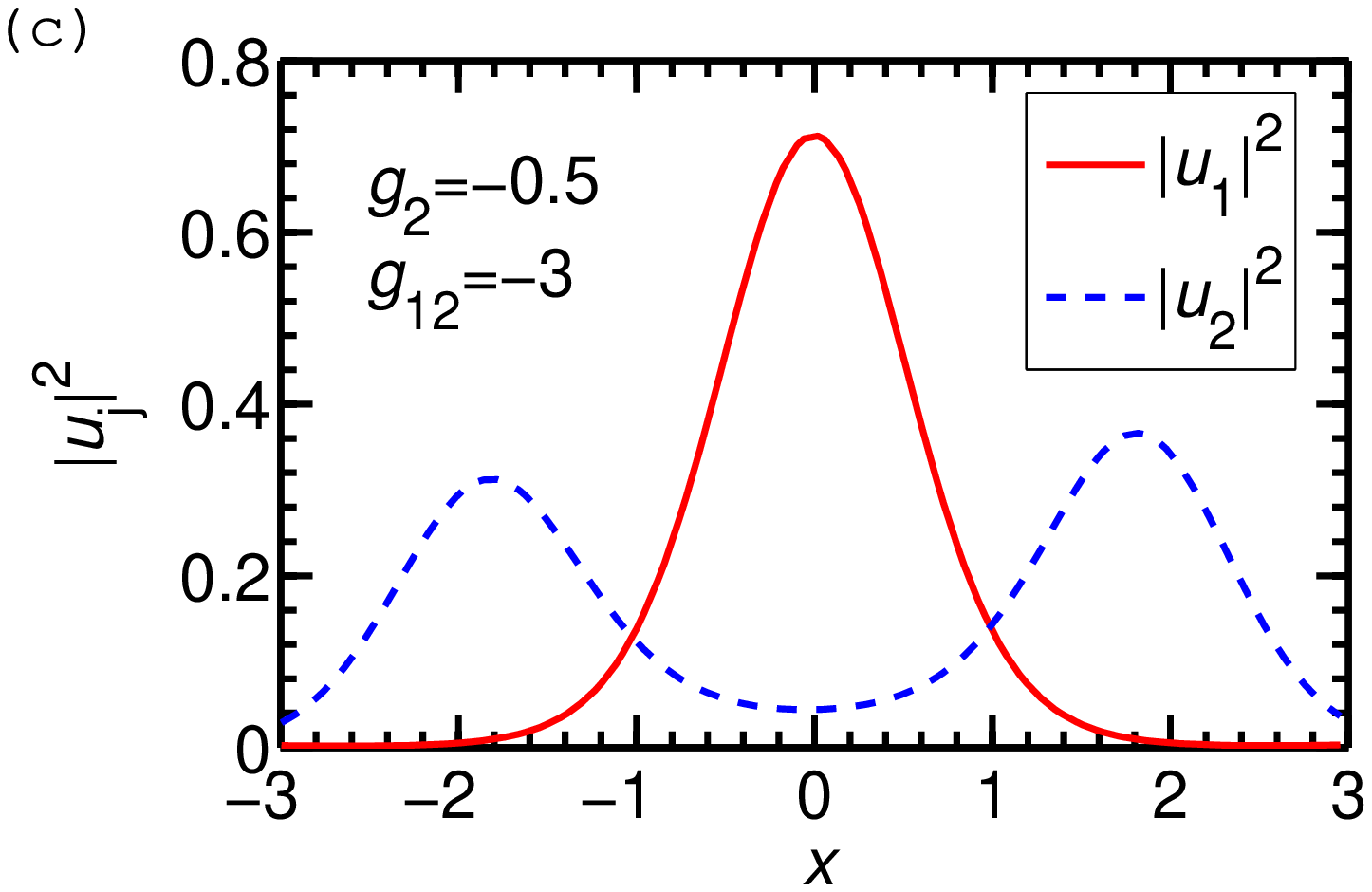}
\includegraphics[width=.49\linewidth]{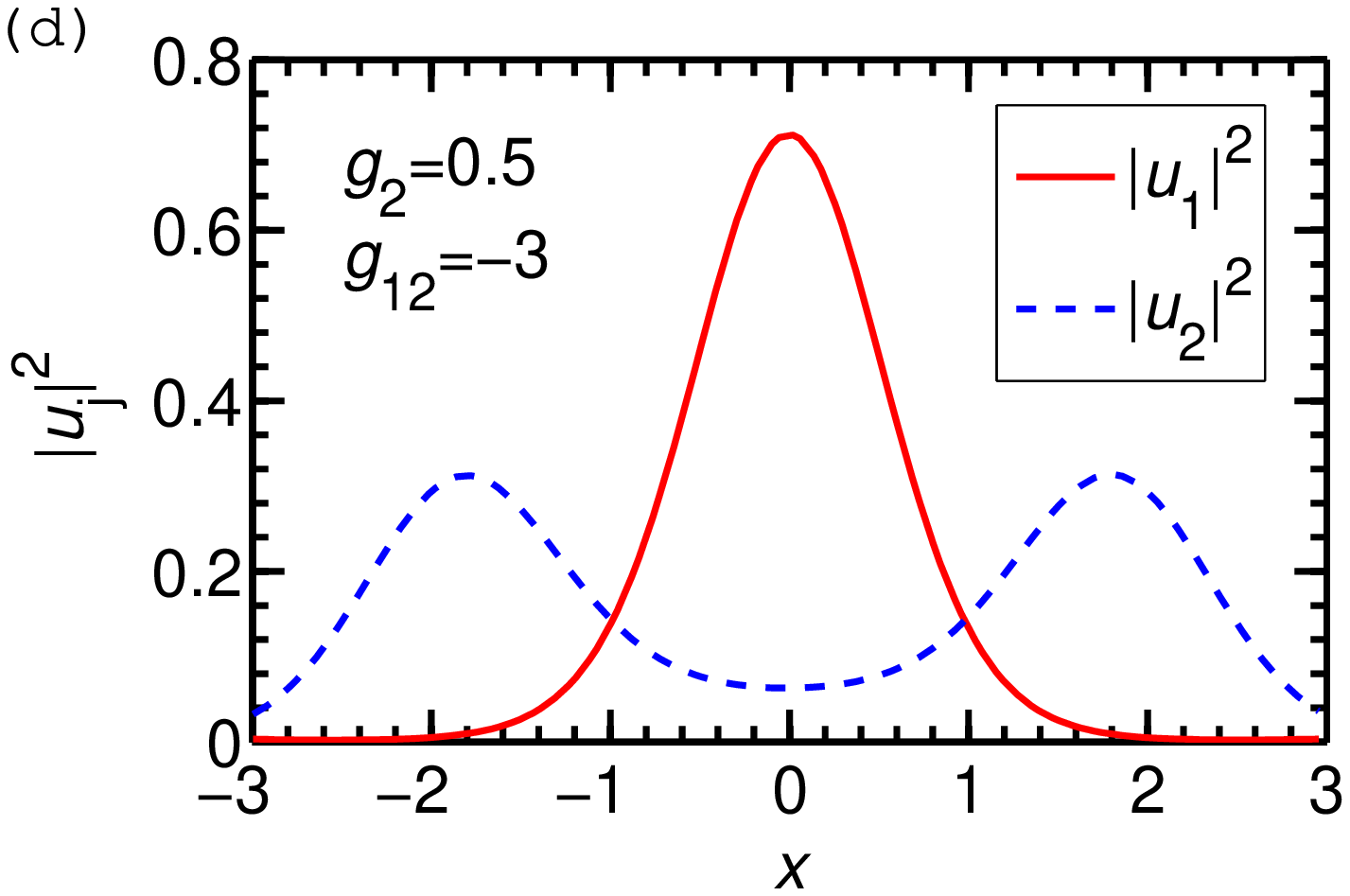}
(c)  \hskip 3 cm   (d)
\end{center}

\caption{(Color online)  The densities $|u_1(x)|^2$ and $|u_2(x)|^2$
of the two localized BEC states 
vs. $x$  for (a)  $g_1=g_2=g_{12}=0,$ (b)  $g_1=g_2=0, g_{12}=-4$,
(c) $g_1=0$, $g_2=-0.5, g_{12}=-3$, and (d) $g_1=0$, $g_2=0.5, g_{12}=-3$, 
when potential (\ref{pot1}) acts on the first component and potential 
(\ref{pot2}) acts on the second component. The potential parameters are 
$\lambda_1=8, \lambda_2=0.862\lambda_1, s_1=8, s_2=2.4.$  
}
\label{prob3}
\end{figure}

Now we consider the situation when potential (\ref{pot1}) acts on component 1
and (\ref{pot2}) acts on component 2.
In this case only the first component under the influence of potential 
(\ref{pot1}) could have a maximum at the origin ($x=0$), and second 
component under the influence of potential (\ref{pot2}) usually has a 
minimum at the center and hence the variational approximation is not 
applicable in this case. {{(The monochromatic counterpart to this
model  with the phase shifted potential (\ref{pot2})
was investigated in Ref. \cite{referee}.)}}
We start our discussion in this case by showing 
results of density under different situations. In Fig. \ref{prob3} (a) 
we plot the densities of the two components for the symmetric 
non-interacting case for $g_1=g_2=g_{12}=0$. The density $|u_1|^2$ of the 
first component has a localized Gaussian-type shape centered at $x=0$. 
The density of the second component has a two-hump symmetric structure 
with a minimum at $x=0$. The symmetry of the densities around $x=0$ can 
be broken in the presence of a sufficiently strong inter-species 
attraction. This is illustrated in Fig.  \ref{prob3} (b) where we plot 
the densities for $g_1=g_2=0$ and $g_{12}=-4$. The center of both the 
densities  has moved slightly towards positive $x$ as a
sufficiently strong  
inter-species attraction  is introduced via $g_{12}=-4$.
In Figs. \ref{prob3} (c) and (d) we illustrate the densities 
in two more cases for $g_1=0, g_2=-0.5, $ and $g_{12}=-3$ and 
 for $g_1=0, g_2=0.5, $ and $g_{12}=-3$, respectively. In the case of 
Fig.  \ref{prob3} (c), the densities are asymmetric and in the case of 
Fig.  \ref{prob3} (d), they are symmetric. This observation is 
consistent with the general conclusion that the symmetry breaking will 
be favored when the system is attractive. In the case of Fig.  
\ref{prob3} (c) we have $g_1=0, g_2=-0.5, $ and $g_{12}=-3$ indicating 
that both components could be bound due to inter-species attraction even 
in the absence of the bi-chromatic OL potential. Hence due to the 
inter-species attraction the two localized BEC states 
tend to overlap and stay together as two bright solitons \cite{cpld}. 
In the case represented in Fig. \ref{prob3} (d) $g_2$ is positive and 
the system is less attractive and hence the density distribution is symmetric.

\begin{figure}
\begin{center}
\includegraphics[width=\linewidth]{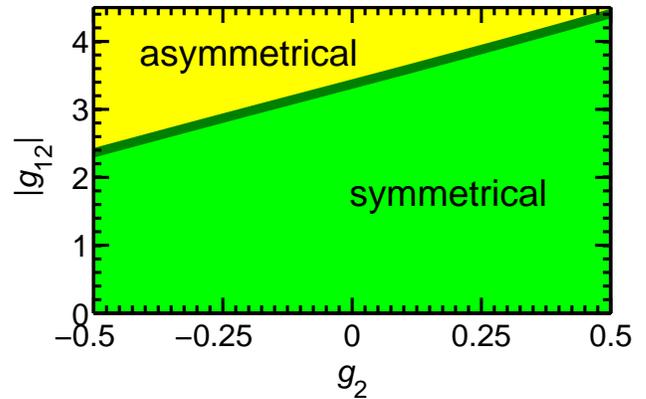}
\end{center}

\caption{(Color online) The phase diagram of  $|g_{12}|$ vs. $g_2$  for $g_1=0$
showing the symmetric  and asymmetric configurations of the 
two localized BEC states. The potential (\ref{pot1}) is
effective on the first component and potential (\ref{pot2}) on the second 
component 
with the same parameters as in Fig.
\ref{phase1}.
}
\label{phase2}
\end{figure}

Again it is worthwhile to show the symmetry breaking in a $g_2$ vs. 
$|g_{12}|$ phase diagram for $g_1=0$ as illustrated in Fig. 
\ref{phase2}. 
{{(Similar symmetry breaking has also been studied 
in the context of symbiotic gap solitons \cite{malomed}.)}}
For small $|g_{12}|$, as expected, the localized states 
are in a symmetric configuration. The asymmetric configuration is 
achieved when $|g_{12}|$ is larger than a critical value indicating that 
a minimum of attraction is needed to replace the center of the localized 
states from $x=0$. 
Symmetry breaking is favored when the inter- and
intra-species interactions
are strongly attractive leading to strong effective
attraction in the system
and hence favored
for large 
negative (attractive)
values of  $g_2$ than for  positive (repulsive) values of  $g_2$ as seen
in Fig. \ref{phase2}.

\begin{figure}
\begin{center}
\includegraphics[width=.49\linewidth]{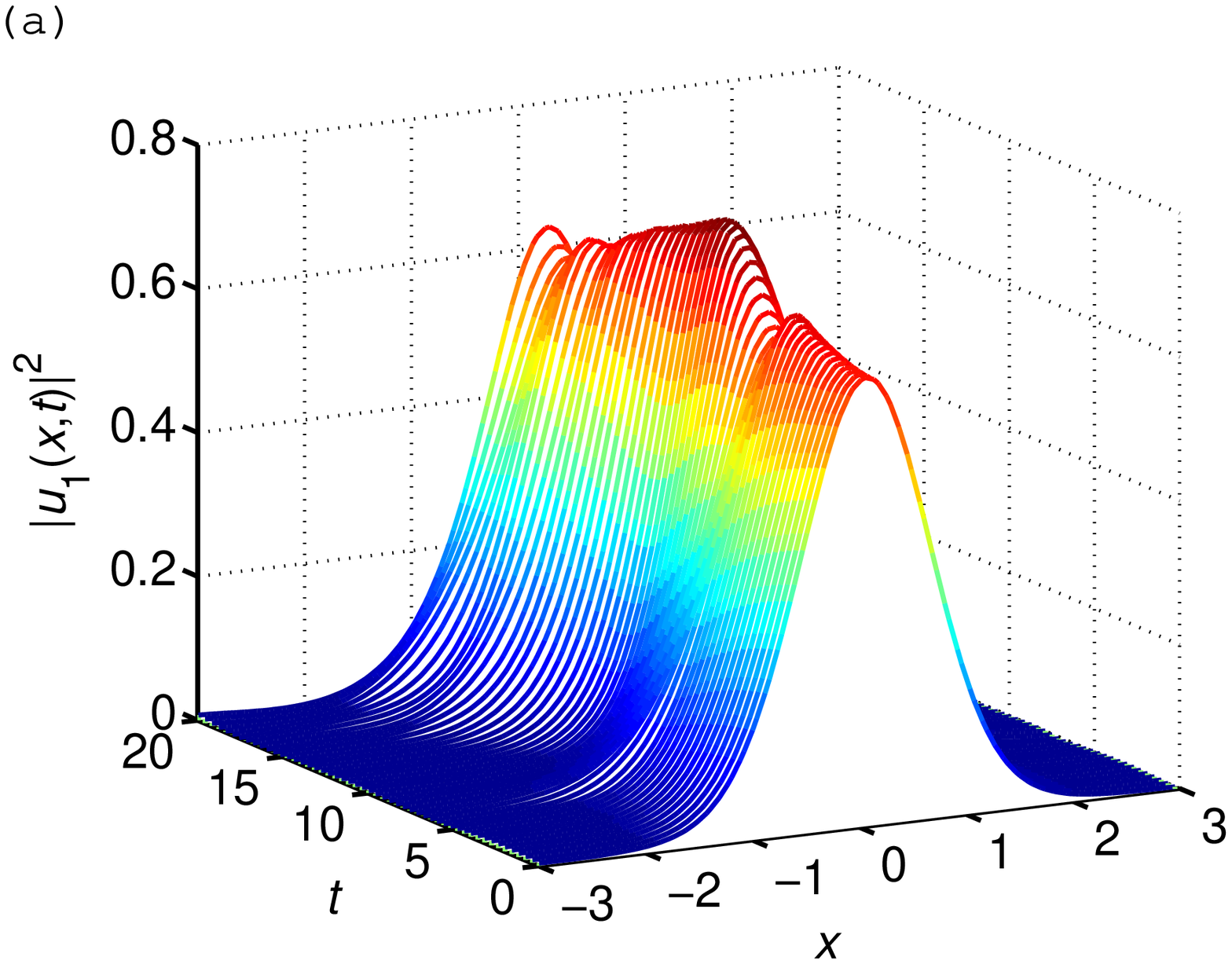}
\includegraphics[width=.49\linewidth]{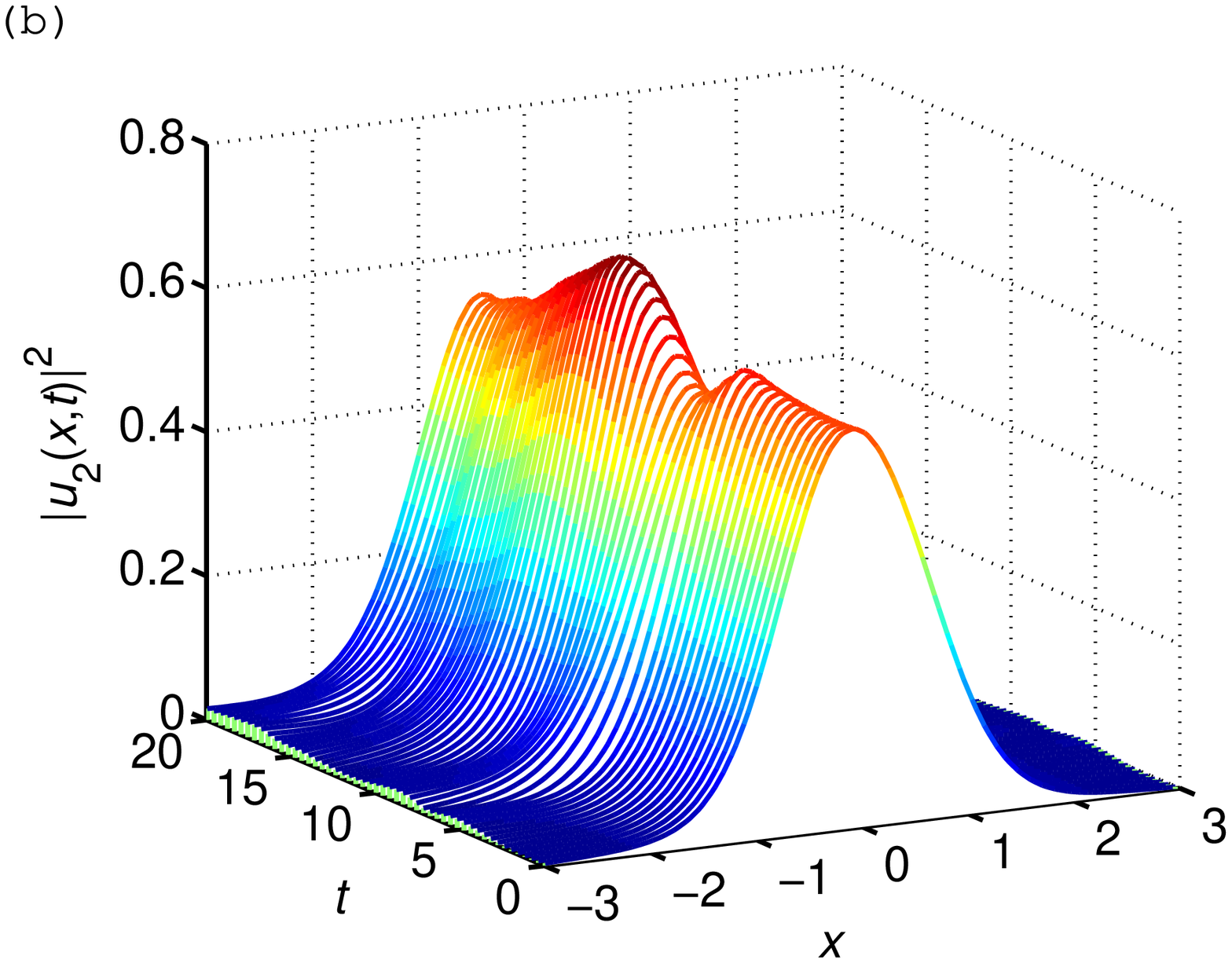}
(a)  \hskip 3 cm   (b)
\end{center}

\caption{(Color online) The densities (a)  $|u_1(x,t)|^2$ and (b) $|u_2(x,t)|^2$ 
vs. $x$ and $t$ of the two localized BEC states of Fig. \ref{prob2} (b) 
when the potential parameters are
suddenly changed from 
$\lambda_1=8, \lambda_2=0.862\lambda_1, s_1=8, s_2=2.4$  to   
$\lambda_1=8.5, \lambda_2=0.862\lambda_1, s_1=8, s_2=2.4$  at time $t=5$.  
}
\label{dyna}
\end{figure}

\subsection{The stability of the localized BEC states}   

\label{IIIC}

So far we studied the stationary properties of the localized BEC
states. Now we investigate if the localized BEC states are stable under 
small perturbation. We are using real-time propagation routines in the 
numerical calculation, which finds usually the stable states. This indicates 
that the localized BEC states, we are   studying, should be stable. However, 
we now explicitly demonstrate the stability of a set of specific states, e.g., 
the ones illustrated in Fig. \ref{prob2} (b). To demonstrate that these
states are stable, after their formation, we suddenly change the potential 
parameters from $\lambda_1=8, \lambda_2=0.862\lambda_1, s_1 =8,  s_2=2.4$ to
 $\lambda_1=8.5, \lambda_2=0.862\lambda_1, s_1 =8, s_2=2.4$ and study 
the resulting dynamics. We plot in Figs. \ref{dyna} (a) and (b) the 
consequent density dynamics $|u_1(x,t)|^2$ and $|u_2(x,t)|^2$ vs. $x$ 
and $t$, respectively, for $t=0$ to 20. The sudden change of the potential 
is effected at $t=5$. From Figs. \ref{dyna} (a) and (b) we see that at 
$t=5$ when the potential is suddenly changed the density profiles suffer 
an abrupt change. Nevertheless, the BEC density profiles remain localized 
for a large interval of time undergoing breathing oscillation as we can see 
from  Figs. \ref{dyna} (a) and (b) for the two components, which demonstrates
the stability of the localized BEC states.

\section{SUMMARY}

In this paper, using the numerical and variational solution of the 
time-dependent GP equation, we studied the localization of two-component 
cigar-shaped BEC in a quasi-periodic 1D OL potential prepared by two 
overlapping polarized standing-wave laser beams with different 
wavelengths and amplitudes. Specifically, we considered two analytical 
forms (sine and cosine) of the OL potential and considered two distinct 
cases: (i) the two components under the action of the sine potential 
(\ref{pot1})
and 
(ii) the sine potential (\ref{pot1})
acting on component 1 and the cosine potential (\ref{pot2}) acting on 
component 2. In case (i), for a weak inter-species interaction, 
both components have a maximum of density at 
the trap center ($x=0$) satisfying the symmetry $|u_j(x)|^2
=|u_j(-x)|^2$. In this case 
 the variational analysis is applicable and 
produced result in good agreement with numerical simulation. In case 
(ii), for a weak inter-species interaction,  
one component has a maximum of density at the trap center and the 
other component has a minimum, however, 
 again satisfying the symmetry $|u_j(x)|^2
=|u_j(-x)|^2$. The localization is most favored for non-interacting atoms and 
is destroyed in the presence of moderate intra-species and inter-species
repulsive  interactions. 
Here we specially study the effect of weak  
inter-species and intra-species nonlinearities on the profile of the 
localized states. In case (i), for a sufficiently strong inter-species 
repulsion, the two localized states are found to move in opposite directions 
from the trap center and attain equilibrium in a separated (split)
configuration breaking the symmetry  $|u_j(x)|^2
=|u_j(-x)|^2$. In case (ii), for a sufficiently strong inter-species 
attraction,  the two localized states are found to move away from $x=0$  
to the same side of the trap center again breaking the symmetry  $|u_j(x)|^2
=|u_j(-x)|^2$.  We studied in some detail the formation of the 
localized states of broken symmetry in all cases considering phase diagram 
of intra- and inter-species nonlinearities $g_2$ and $g_{12}$. 

{{The Anderson localization of a BEC in a bi-chromatic OL 
potential as studied here is very distinct from the localization of gap 
solitons \cite{gsexp,gs,referee} of the BEC in a mono-chromatic OL potential. Anderson 
localization takes place in an aperiodic potential in the linear 
equation, whereas a periodic  potential and a repulsive nonlinearity 
are crucial for the localization of gap solitons.}  
}

We hope that the present work will motivate new studies, specially 
experimental ones, on the localization of a binary mixture of BEC in a 
bi-chromatic OL potential. The effect of nonlinear inter- and intra-species 
interactions on localization, as investigated in the 
present paper,  also deserves careful analysis.  

\label{IIII}

\acknowledgments

FAPESP (Brazil) CNPq (Brazil) 
provided partial support. YC undertook this work with the support 
of the post-doctoral  program of FAPESP (Brazil).


\end{document}